\documentclass[10pt,aps,twocolumn,prd,superscriptaddress,noshowpacs,nofootinbib,noshowkeys,floatfix]{revtex4-1}
\usepackage{graphicx}

\usepackage{multirow}
\usepackage{longtable}
\usepackage[normalem]{ulem}

\usepackage{booktabs}
\usepackage[hidelinks]{hyperref}
\usepackage{color}

\renewcommand\sout{\bgroup \color{blue} \ULdepth=-.5ex \ULset}
\hypersetup{
    colorlinks,
    linkcolor={blue},
    citecolor={blue},
    urlcolor={blue}
}
\usepackage{amssymb,amsmath,amsfonts}
\usepackage[utf8]{inputenc}
\usepackage{nicefrac}
\usepackage{upgreek}
\usepackage{braket}
\usepackage{slashed}

\begin{document}

\title{ Fourier coefficients of the net baryon number density and chiral criticality
 }

\author{G\'abor Andr\'as Alm\'asi}
\affiliation{Wigner Research Center for Physics, H-1121 Budapest, Hungary}
\author{Bengt Friman}
\affiliation{GSI Helmholtzzentrum f\"ur Schwerionenforschung GmbH,
D-64291 Darmstadt, Germany}
\author{Kenji Morita}
\affiliation{Institute of Theoretical Physics, University of Wroc{\l}aw, PL-50204, Wroc{\l}aw, Poland}
\affiliation{iTHES Research Group, RIKEN, Saitama 351-0198, Japan}
\author{Pok Man Lo}
\affiliation{Institute of Theoretical Physics, University of Wroc{\l}aw, PL-50204, Wroc{\l}aw, Poland}
\author{Krzysztof Redlich}
\affiliation{Institute of Theoretical Physics, University of Wroc{\l}aw, PL-50204, Wroc{\l}aw, Poland}
\affiliation{Research Division and EMMI, GSI Helmholtzzentrum f\"ur Schwerionenforschung,
\\ 64291 Darmstadt, Germany}

\date{\today}

\begin{abstract}
 We investigate  the Fourier coefficients $b_k(T)$ of the  net baryon
 number density in strongly interacting matter at nonzero temperature and density.
 The asymptotic behavior of the coefficients at large $k$ is determined
 by the singularities of the partition function in the complex chemical potential plane.
 Within a QCD-like effective chiral model, we show that the chiral and deconfinement properties at
 nonzero baryon chemical potential are reflected in characteristic
 $k$- and $T$- dependencies of the Fourier coefficients. We also discuss the influence of the
 Roberge-Weiss transition on these coefficients. Our results indicate that
 the Fourier expansion approach can provide interesting insights into the
 criticality of QCD matter.
\end{abstract}


\maketitle
\section{Introduction \label{sec:Introduction}}

Exploring the phase diagram of quantum chromodynamics (QCD) at finite
temperature and chemical potentials is a challenging problem in
experimental and theoretical studies.
The Beam Energy Scan (BES) at the Relativistic
Heavy Ion Collider (RHIC) \cite{luo16:_explor_qcd} has been dedicated to
the search for the conjectured  QCD critical point (CP) and the onset of
deconfinement through the systematic studies of various observables
such as the fluctuations of conserved charges. In particular, those of
the net proton number have been shown to exhibit a nonmonotonic
behavior with collision energies which may potentially be attributed
to critical chiral dynamics
\cite{Ejiri:2005wq,karsch:QM17,Bazavov:2017tot,skokov11:_quark_number_fluct_in_polyak,Karsch:2010hm,stephanov09:_non_gauss_fluct_near_qcd_critic_point,stephanov11:_sign_of_kurtos_near_qcd_critic_point,Almasi:2017bhq}.

Fluctuation observables can also be computed in lattice QCD (LQCD)
\cite{bazavov12:_fluct_and_correl_of_net,borsanyi12:_fluct_of_conser_charg_at,Ichihara:2015kba}
and in effective chiral models
\cite{Karsch:2010hm,sasaki07:_quark_number_fluct_in_chiral,skokov10:_meson_fluct_and_therm_of,friman11:_fluct_as_probe_of_qcd,Almasi:2017bhq}.
Thus, by identifying fluctuation patterns, associated with chiral critical behavior, in experimental data,
one may be able to identify the QCD phase boundary and the location of the CP, if it exists \cite{Ejiri:2005wq,Karsch:2010hm,Almasi:2017bhq,Almasi:2016gcn,Asakawa:2015ybt}.

While LQCD provides first principle insights into the thermodynamics
of QCD at finite temperature $T$ and at vanishing and moderate values of
the  baryon chemical potential $\mu_B<3T$, the well-known sign problem still
inhibits systematic calculations at larger baryon densities. Several strategies to
overcome this problem are being pursued \cite{muroya03:_lattic_qcd,forcrand09:_simul_qcd_at_finit_densit}.

In particular, LQCD calculations at imaginary baryon chemical potential offer a
possibility to circumvent the sign problem at real $\mu_B$. On the one
hand, these calculations, which do not suffer from the sign problem, can,
in principle, be analytically continued to the real axis
\cite{forcrand02:_qcd,forcrand03:_qcd,d'elia03:_finit_qcd}.
On the other, a Fourier expansion in imaginary $\mu_B$ can
be applied to the partition function in order to study the properties, in particular the phase
structure, of QCD at finite baryon density in the canonical ensemble
\cite{ejiri08:_canon_qcd,nagata14:_lee_yang_qcd_rober_weiss,Boyda:2017dyo,Bornyakov:2017wzr}.
Moreover, the canonical partition function can be used to obtain the probability
distribution function $P(N)$ of the net baryon number $N$,  which in turn yields the cumulants of the net baryon number fluctuations
\cite{braun-munzinger11:_net_proton_probab_distr_in,braun-munzinger12:_net_charg_probab_distr_in,morita12:_baryon_number_probab_distr_near,morita13:_net,morita14:_critic_net_baryon_number_probab,Nakamura:2013ska,Morita:2015tma}. 
As noted in Ref.~\cite{morita14:_critic_net_baryon_number_probab}, knowledge of the probability for
large-amplitude fluctuations, i.e., fluctuations with large $N$, is required for a correct identification of the critical properties associated with the chiral phase transition.

Furthermore, a novel strategy for locating the confinement-deconfinement
transition by exploring the complex phase of the Polyakov loop at imaginary
chemical potential has been proposed in
\cite{Kashiwa:2016vrl,Kashiwa:2017swa}. Thus, by studying
QCD at imaginary chemical potential, one can gain insight into the
critical properties of QCD related to deconfinement, chiral
symmetry restoration and the Roberge-Weiss transition.

In this paper we explore the Fourier decomposition of the (dimensionless) net baryon number density
\begin{equation}\label{eq:susc1}
\chi_1^B(T,\hat{\mu}_B)\equiv \frac{n_B(T,\hat{\mu}_B)}{T^3}=\frac{\partial (p/T^4)}{\partial \hat{\mu}_B },
\end{equation}
where $p=p(T,\hat{\mu}_B)$ is the pressure, $\hat{\mu}_B=\mu_B/T$
is the reduced baryon chemical potential and $T$ the temperature.
The analytic continuation of $\chi_1^B$ to imaginary baryon chemical potential
is an odd, periodic function of   $\theta_B=\text{Im}\,\hat{\mu}_B$, and can thus be expanded in
a Fourier sine series
\begin{equation}
\label{eq:imagdens}
\textrm{Im}[
\chi_1^B(T,i\,\theta_B)] = \sum_{k=1}^{\infty} b_k(T) \sin(k\,\theta_B).
\end{equation}
The aim of our study is to
identify the influence of chiral symmetry restoration and deconfinement
on the Fourier coefficients $b_k(T)$.
We examine the effects of criticality by making use of QCD-like chiral
effective models. In particular, we discuss the effect of singularities
in the complex chemical potential plane
associated with first- and second-order phase transitions,
crossover transitions and the contribution of the so-called
thermal singularities. Model-independent results will be
reported in a separate publication~\cite{Fourier_general}.

To be specific, we employ the two-flavor Polyakov-quark-meson (PQM)
model, which emulates the characteristic
properties of QCD  both at real~\cite{schaefer07:_phase_polyak} and
imaginary~\cite{morita11:_role_of_meson_fluct_in} baryon chemical
potentials. At nonzero $T$
and $\theta_q = \theta_B/N_c$, the resulting  thermodynamic potential exhibits
chiral symmetry restoration and statistical confinement, as well as the
Roberge-Weiss symmetry~\cite{roberge86:_gauge_qcd}, which implies a periodicity
of $2\pi/N_c$ in $\theta_q$.

We show that the $k$- and $T$- dependencies of the Fourier coefficients
$b_k(T)$ exhibit characteristic features, reflecting the chiral and  Roberge-Weiss transitions at imaginary baryon chemical potential.

The results are discussed in light of the Fourier
expansion coefficients that were recently obtained in LQCD simulations at imaginary chemical potential for a wide range of temperatures around and above the chiral and deconfinement transitions~\cite{Vovchenko:2017xad}. Furthermore, we
examine the recently proposed cluster expansion model
\cite{Vovchenko:2017gkg} for Fourier coefficients and discuss its
predictive power and analytic properties in the context of chiral
criticality.

The paper is organized as follows: In the next section, we analyze
phenomenological models for the Fourier coefficients and their
applicability to the description of criticality and interpretation of recent LQCD results. The properties of the
Fourier expansion coefficients in a QCD-like effective chiral model are examined in Sec.~\ref{sec:PQM}. Finally, section \ref{sec:summary} is devoted to a summary and conclusions.

\section{Modeling the Fourier coefficients and lattice QCD\label{sec:Modeling}}

\subsection{Models for Fourier coefficients}

A model for the Fourier coefficients of the net baryon density, the cluster expansion model (CEM), was recently proposed in
Ref.~\cite{Vovchenko:2017gkg}. Based on the popular hadron-resonance gas equation of state with excluded-volume corrections, the authors suggested a simple prescription for computing the higher-order Fourier coefficients of the net baryon
density $b_k(T)$ in Eq.~\eqref{eq:imagdens} in terms of the first two, $b_1(T)$ and $b_2(T)$,
\begin{equation}
 \label{eq:CEMdef}
   b_k^{\text{CEM}}(T) =  \left(\frac{b_1^{SB}}{b_1(T)}\right)^{k-2}\left(\frac{b_2(T)}{b_2^{SB}}\right)^{k-1}\,b_k^{SB},
\end{equation}
where
\begin{equation}\label{eq:bk_SB}
 b_k^{SB} = (-1)^{k-1} \frac{ 12+16(\pi k)^2}{27\, k \, (\pi k)^2}
\end{equation}
are the Fourier coefficients of the density of a noninteracting gas of massless quarks with $N_c=3$ colors and $N_f=3$ flavors.
The coefficients (\ref{eq:CEMdef}) are constructed such that they all approach the corresponding Stefan-Boltzmann (SB) values  (\ref{eq:bk_SB}), when $b_1(T)$ and $b_2(T)$ approach $b_1^{SB}$ and $b_2^{SB}$, respectively.

The first four Fourier coefficients, $b_1(T) - b_4(T)$, have been
obtained in LQCD calculations~\cite{Vovchenko:2017xad}. It was
shown in Ref.~\cite{Vovchenko:2017gkg} that Eq.~\eqref{eq:CEMdef},
using the lattice results as input for $b_1(T)$ and $b_2(T)$,
provides a  description of $b_3(T)$ and $b_4(T)$ consistent with
LQCD. Furthermore, the model predictions for the temperature dependence
of the sixth-order baryon number susceptibility are qualitatively
consistent with LQCD findings.

Thus, the CEM describes the basic features of the available lattice results on baryon number fluctuations. Nevertheless, one may ask whether the critical behavior of QCD associated with the restoration of chiral symmetry, which is reflected in the asymptotic behavior of the Fourier
coefficients, can be captured by the model. More generally, it is of interest to assess to which extent the modeling of Fourier coefficients is unique, when only the first two coefficients are provided as input.

\begin{figure*}[!tb]
 \centering
 \includegraphics[width=0.49\linewidth]{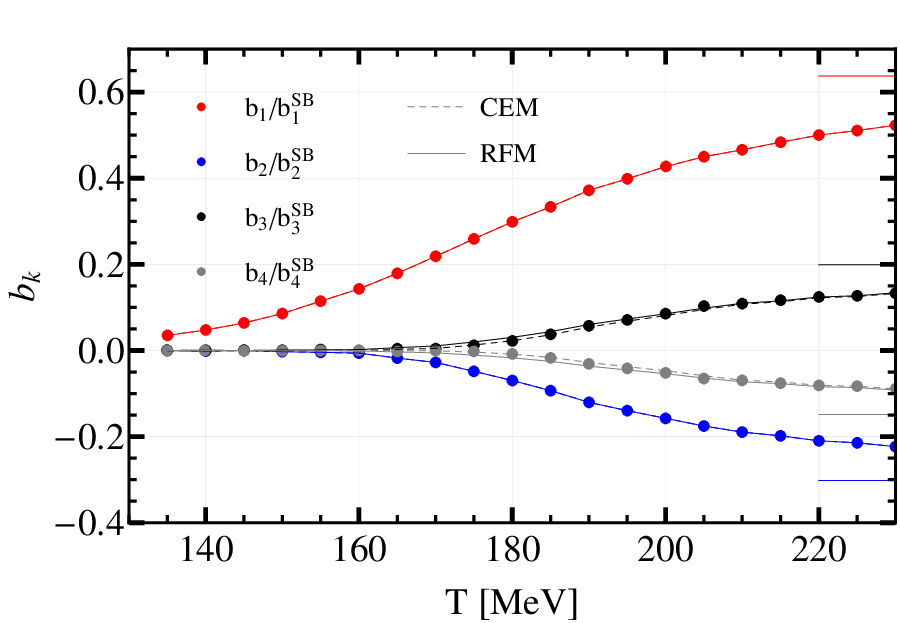}
 \includegraphics[width=0.49\linewidth]{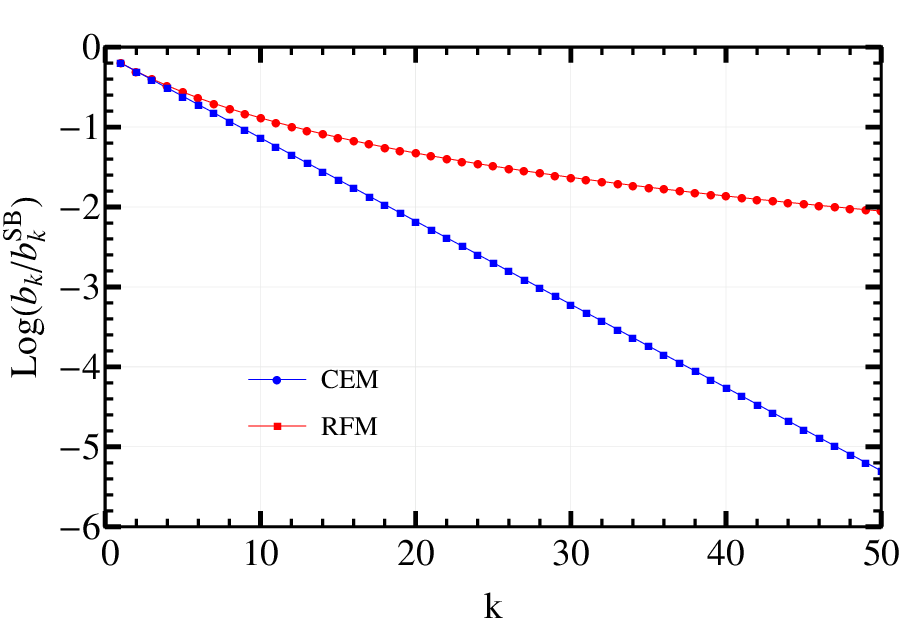}
 \caption{Left: Temperature dependence of the Fourier coefficients in
 the cluster expansion model (CEM)  \cite{Vovchenko:2017gkg} and the
 rational function model (RFM), defined in  Eq.~\eqref{ref1}, compared to LQCD
 data~\cite{Vovchenko:2017xad}. Right: The Fourier
 expansion coefficients obtained in these  models
 at $T={230\;\mathrm{MeV}}$. \label{fig:coeffs}}
\end{figure*}

In order to illustrate these issues, 
we introduce an alternative model, which we dub the ``rational fraction model''
(RFM). The functional form of the RFM Fourier coefficients,\footnote{Note that the prefactor in \eqref{ref1} is an even function of $k$, and thus preserves the symmetry of the sine Fourier coefficients \eqref{eq:bk_SB}, i.e., $b^{\text{RFM}}_{-k}=-b^{\text{RFM}}_{k}$.}
\begin{equation}
\label{ref1}
b_k^{\text{RFM}}(T) = \frac{c(T)}{1+\sqrt{k^2}/k_0 (T)} b_k^{SB},
\end{equation}
with
\begin{align}
k_0(T) &=\left[ \frac{b_1(T) b_2^{SB}}{b_2(T)b_1^{SB}}-1 \right]^{-1}-1, \label{ref2}\\
c(T) &= \frac{b_1(T)}{b_1^{SB}}\left(1+\frac{1}{k_0(T)}\right),\label{ref3}
\end{align}
was chosen so that the asymptotics is a power law and the SB limit as well as the lattice results on Fourier coefficients are reproduced. By contrast, the CEM Fourier coefficients (\ref{eq:CEMdef}) fall off exponentially for $k\to \infty$. 
In the left panel of Fig.~\ref{fig:coeffs}, we show the temperature dependence of the CEM and RFM Fourier coefficients $b_1\dots b_4$. The overall agreement with the LQCD results~\cite{Vovchenko:2017xad} is of similar quality for the two models.

\begin{figure}[!bh]
  \centering
  \includegraphics[width=0.99\linewidth]{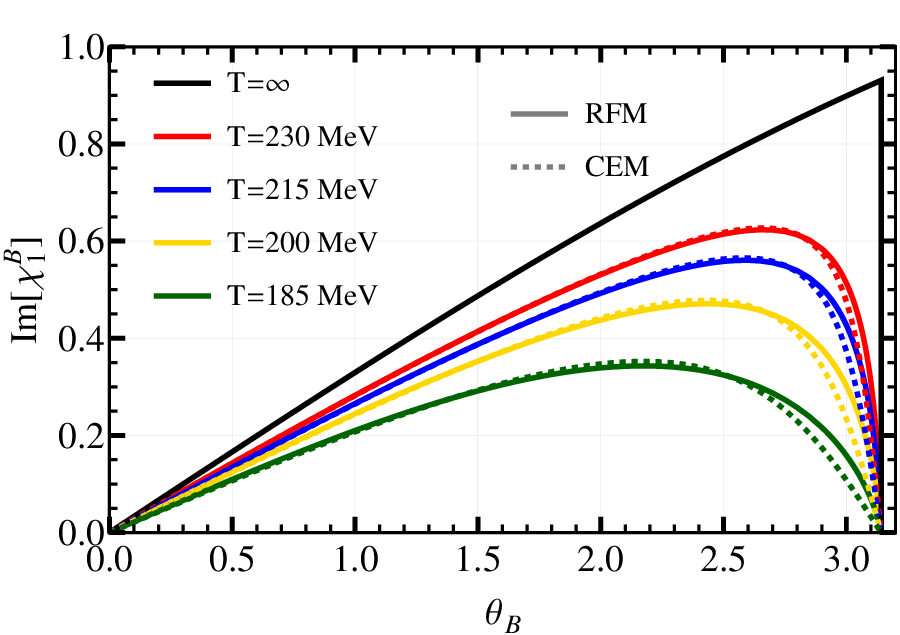}
  \caption{The imaginary part of the baryon density at imaginary baryon chemical
 potential for various temperatures obtained in the rational function
 model (RFM) and cluster expansion model (CEM). \label{fig:dens}}
\end{figure}

The asymptotic behavior of the Fourier coefficients in the CEM and RFM is illustrated in the right panel of Fig.~\ref{fig:coeffs}. 
As discussed below, the change from exponential to power-law asymptotics of the Fourier coefficients is a consequence of a different analytical structure of the density in the complex chemical potential plane and is also reflected in modifications of higher cumulants of the net baryon number. This example illustrates the fact that the asymptotic form of a Fourier series and consequently the analytic structure of the function are not fixed by the first few Fourier coefficients.

For real values of the baryon chemical potential, the net baryon number density (\ref{eq:susc1}) is given by
\begin{equation}
 \label{eq:realdens}
  \chi_1^B(T,\hat{\mu}_B) = \sum_{k=1}^{\infty} b_k(T) \sinh(k\hat{\mu}_B),
\end{equation}
where $\hat{\mu}_B=\mu_B/T$ and $b_k(T)$ are the Fourier coefficients
of the density at imaginary chemical potential (\ref{eq:imagdens}),
\begin{equation}
 b_k(T)= \frac{2}{\pi}\int_0^\pi d\theta_B
  \big[\mathrm{Im}\chi_1^B(T,i\theta_B)\big]\sin(k\theta_B). \label{eq:bk_Fourier}
\end{equation}
Higher-order fluctuations of the net baryon number are
obtained by taking derivatives of Eq.~\eqref{eq:realdens},
\begin{equation}
 \chi_n^B = \sum_{k=1}^{\infty} b_k
  \frac{\partial^{n-1}}{\partial\hat{\mu}_B^{n-1}}\sinh(k\hat{\mu}_B).\label{eq:cumulants}
\end{equation}

In Fig.~\ref{fig:dens} we show $\textrm{Im}[\chi_1^B(T,i\,\theta_B)]$, obtained with
Eq.~\eqref{eq:imagdens} using the Fourier coefficients
\eqref{eq:CEMdef} and \eqref{ref1} at several temperatures.
Since the first four coefficients coincide in CEM and RFM, the
difference between the two models for the baryon number density at
imaginary chemical potential is relatively small and discernible only at $\theta_B \simeq
\pi$. There, by construction, both functions drop rapidly to zero and
the convergence of the Fourier sum is slow.

Differences between the higher-order Fourier coefficients imply very
different predictions for the baryon number susceptibilities $\chi_n^B$, in particular
at large $n$.

\begin{figure*}[ht!]
\centering
 \includegraphics[width=0.49\linewidth]{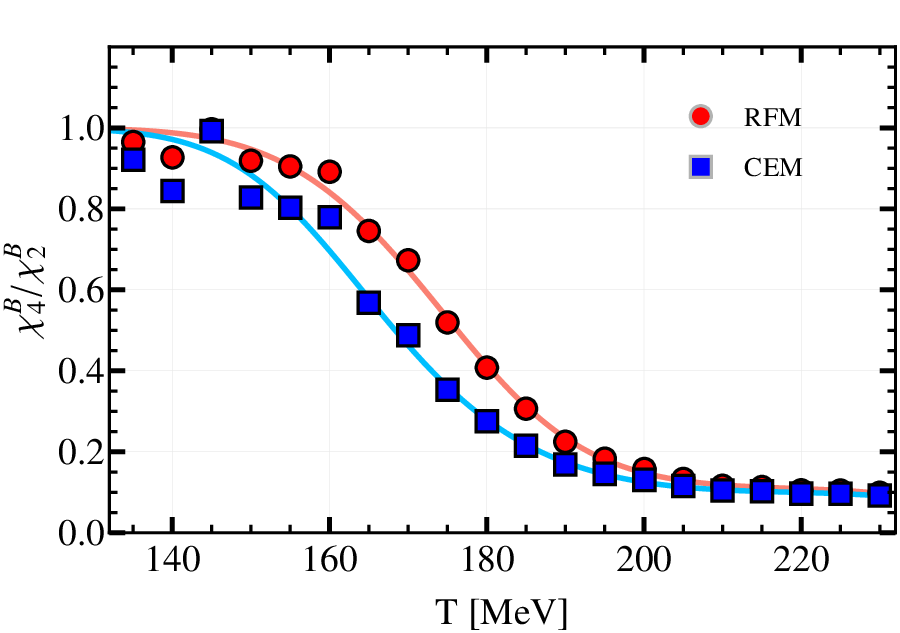}
 \includegraphics[width=0.49\linewidth]{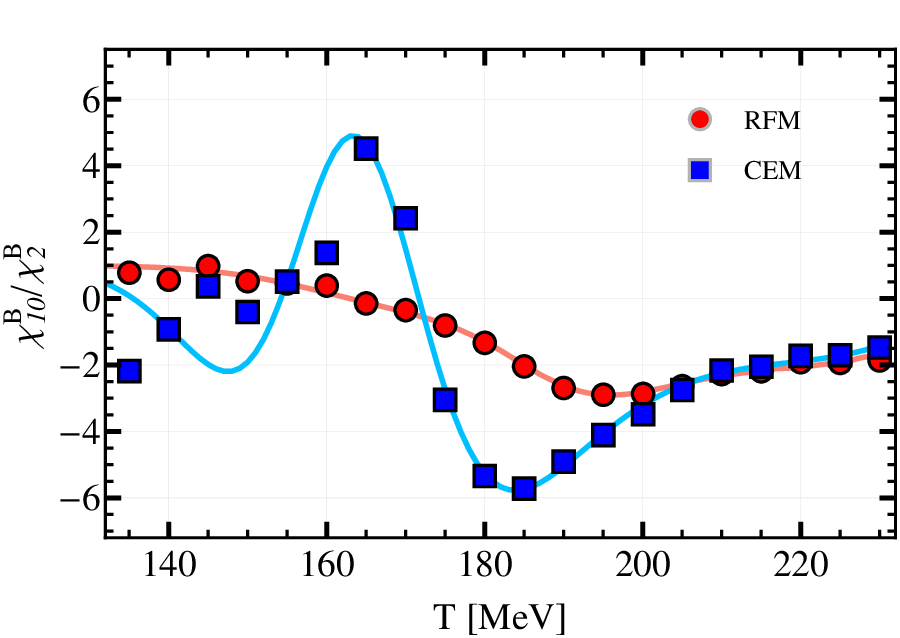}
 \caption{Baryon number cumulant ratios obtained in the  cluster
 expansion model  \cite{Vovchenko:2017gkg} and the rational function model  of  Eq.~\eqref{ref1}. The  points are computed directly using the LQCD data points~\cite{Vovchenko:2017xad} for $b_1$ and
 $b_2$ and the two models for the  higher-order Fourier
 coefficients. The solid curves were obtained by fitting $b_1$ and $b_2$ to the LQCD data, as described in the text.  \label{fig:cumulants}}
\end{figure*}

In Fig.~\ref{fig:cumulants} we show the fourth- and tenth-order
cumulants normalized by $\chi_2^B$ in CEM and RFM. Each point represents
the result of a model calculation, where the Fourier coefficients were obtained using Eqs. (\ref{eq:CEMdef}) and (\ref{eq:bk_SB}) and (\ref{ref1})-(\ref{ref3}), respectively, while $b_1(T)$ and $b_2(T)$ are given by LQCD data.
The lines in Fig.~\ref{fig:cumulants} are obtained
by interpolating the LQCD values for $b_1(T)$ and $b_2(T)$ as functions of temperature in the range
$T\in[165\;\textrm{MeV},220\;\textrm{MeV}]$. We thus obtain the following fit to the LQCD data
\begin{align}\label{eq:fit-Fourier}
	\log (b_1) &= -41.5 + 83.1 x - 56.8 x^2 + 13.1 x^3, \nonumber \\
	\log (-b_2) &= -135.4 + 280.2 x - 196.4 x^2 + 46.1 x^3,
\end{align}
with $x=T/155~\text{MeV}$.

{As shown in the left panel of Fig.~\ref{fig:cumulants}, the two models yield similar behavior for the temperature dependence of  the   $\chi_4^B/\chi_2^B$ ratio. However, higher-order cumulants are very  different in these models
due to  the contribution of higher-order Fourier coefficients.
This is illustrated
in the right panel of Fig.~\ref{fig:cumulants}  where we show that the temperature dependence of the $\chi_{10}^B/\chi_2^B$  ratio in CEM and RFM   differs essentially
 in the crossover region.
 This clearly demonstrates that the characteristic features of higher-order fluctuations, which are sensitive to criticality, are not uniquely determined by requiring that the first four Fourier coefficients be reproduced. Thus, in contrast to Ref.~\cite{Vovchenko:2017gkg}, we find  that models of this type are in general  not suited for exploring chiral criticality.

 In  fact,  in the following   we show   that for the phenomenological
 CEM  the Fourier series in Eq. (\ref{eq:imagdens}) can be resummed and expressed by the polylogarithm functions which
have well-defined analytic structure, excluding explicitly any information
on  criticality expected in QCD. This is also the case for the RFM.}

\subsection{Analytic properties and criticality \label{sec:analytic}}

Given the Fourier coefficients of the net baryon density, one can compute susceptibilities and consequently obtain information on the location of singularities in the complex $\mu_B$ plane, e.g., by computing the radius of convergence of the Taylor expansion of the pressure. In order to assess whether the CEM exhibits any singularities that could provide insight into the critical behavior of QCD, we examine the analytical properties of the model. Of particular interest in this context are the questions concerning the location of singularities associated with the  QCD chiral transition.

We first consider the Fourier expansion of the density in the CEM in the
Stefan-Boltzmann (SB) limit.  In this case, the baryon density
is computed using the Fourier coefficients \eqref{eq:bk_SB}
in \eqref{eq:realdens},
\begin{align}
 &\chi_1^{B,SB} =\sum_{k=1}^{\infty} b_k^{SB} \sinh\left(k\hat{\mu}_B\right)
 \label{eq:expsum} \\
 &= \frac{2}{27\pi^2} \sum_{s=\pm 1} s \left[3 \sum_{k=1}^{\infty}
 \frac{\left(- e^{-s\hat{\mu}_B}\right)^k}{k^3} +4\pi^2
 \sum_{k=1}^{\infty} \frac{\left(- e^{-s\hat{\mu}_B}\right)^k}{k}
 \right] \nonumber \\
 &= \frac{2}{27\pi^2} \sum_{s=\pm 1} s \left[ 3 \mathrm{Li}_3(- e^{-s\hat{\mu}_B}) -4\pi^2 \log\left(1 +  e^{-s\hat{\mu}_B} \right) \right],\nonumber
\end{align}
where $\mathrm{Li}_n(z)$ denotes the polylogarithm of order $n$
defined by
\begin{equation}
 \mathrm{Li}_n(z) = \sum_{k=1}^{\infty} \frac{z^k}{k^n}.
\end{equation}
We note that Eq.~\eqref{eq:expsum} is valid also for complex values of the chemical potential. Moreover, for $|\mathrm{Im}\,\hat{\mu}_B|<\pi$, Eq.~\eqref{eq:expsum} reduces to
\begin{equation}
 \label{eq:polynomimalchi}
  \chi_1^{B,SB}=\frac{\hat{\mu}_B}{3}+\frac{\hat{\mu}_B^3}{27\pi^2},
\end{equation}
while for $|\mathrm{Im}\,\hat{\mu}_B|>\pi$, the function is consistent with the periodicity in the imaginary $\hat{\mu}_B$ direction, which is implemented in the model.

Now, the density of an ideal gas of massless quarks and antiquarks
is given by the well-known expression
\begin{align}
\label{eq:SBdens}
 \chi^{B,IG}_1 &= \frac{2N_c N_f}{3} \int \frac{d^3 \hat{p}}{(2\pi)^3}
 \left(\frac{1}{1+e^{\hat{p}-\hat{\mu}_B/3}}
 -\frac{1}{1+e^{\hat{p}+\hat{\mu}_B/3}}\right)
 \nonumber \\
 &= \frac{2 N_c N_f}{3\pi^2}\left( \mathrm{Li}_3\left(
 -e^{-\hat{\mu}_B/3}\right)-\mathrm{Li}_3\left( -e^{\hat{\mu}_B/3}
 \right) \right),
\end{align}
where $\hat{p}=p/T$.
For $|\mathrm{Im}\hat{\mu}_B|< 3\pi$, Eq.~\eqref{eq:SBdens} reduces to the polynomial
form \eqref{eq:polynomimalchi}.
Consequently, $ \chi^{B,SB}_1$ and $\chi_1^{B,IG}$ coincide
within a band in the complex $\hat{\mu}_B$ plane defined by $|\mathrm{Im}\hat{\mu}_B|< \pi$ (as well as in bands obtained by shifting $\mathrm{Im}\,\hat{\mu}_B$ by  multiples of $6\,\pi$).

Outside these bands, the two densities differ, due to the different periodicities: $ \chi^{B,SB}_1$  is invariant under translations of the baryon chemical potential by multiples of $2\,\pi\, i\, T$, i.e., $\hat{\mu}_B \rightarrow \hat{\mu}_B + 2\,\pi\, i\, N$, while $\chi_1^{B,IG}$ is invariant under shifts by multiples of $6\,\pi\, i\, T$, i.e.,
$\hat{\mu}_B \rightarrow \hat{\mu}_B + 6\,\pi\, i\, N$, where $N$ is an arbitrary integer.
The periodicity of $\chi_1^{B,SB}$ is a consequence of the Roberge-Weiss symmetry~\cite{roberge86:_gauge_qcd}, which implies that only states with an integer net baryon number contribute to the partition function. Likewise, the $6\,\pi\, i$ periodicity of $\chi_1^{B,IG}$ is a consequence of the states with fractional baryon number accessible in a gas of noninteracting quarks.

Closely related to the periodicity of $\chi_1^{B,SB}$ and $\chi_1^{B,IG}$ is the location of singularities in the complex $\hat{\mu}_B$ plane. Both $\mathrm{Li}_n(z)$ and  $\log(1-z)$ have branch points at $z=1$. Hence, the singularities of $\chi_1^{B,SB}$ closest to $\hat{\mu}_B=0$ are located  on the imaginary axis, at $\hat{\mu}_B=\pm i\,\pi$.  On the other hand, the closest singularities of the ideal quark gas, $\chi_1^{B,IG}$, are found at $\hat{\mu}_B=\pm 3\, i\,\pi$. We note that the latter is generated by the pole of the Fermi-Dirac function at $\hat{p}=0$. For nonzero quark mass $m$, these thermal branch points are shifted away from the imaginary axis to $\hat{\mu}_B=\pm\, \hat{m} \pm\, 3\,i\,\pi$, where $\hat{m}=m/T$.

Thus, the singularity structure of the net baryon density is, in the Stefan-Boltzmann limit, completely determined by the analytic properties of the  polylogarithm. In the following, we show that this is the case also for the CEM.

We first note that the Fourier coefficients of the baryon density in
Eq.~\eqref{eq:CEMdef} can be expressed in the following form,
\begin{equation}
 \label{eq:CEMdef2}
  b_k^\text{CEM}(T)=
 c \lambda^k b_k^{SB}
\end{equation}
where
\begin{align}
 \label{par}
 c(T) &= \left[ \frac{b_1(T)}{b_1^\text{SB}} \right]^2
 \frac{b_2^\text{SB}}{b_2(T)}, \\
 \lambda(T) &=\label{eq:par2} \frac{b_1^\text{SB}}{b_1(T)}\frac{b_2(T)}{b_2^\text{SB}}.
\end{align}
Also in this case, a closed-form expression for the density can be obtained by resumming the Fourier series,
\begin{align}
 &\chi_1^{B,\text{CEM}}=c\sum_{k=1}^{\infty} \lambda^k b_k^{CEM}
 \sinh\left(k\hat{\mu}_B\right) \label{eq:expsum2} \\
 &= \frac{2 c}{27\pi^2}	\sum_{s=\pm 1} s \left[ \sum_{k=1}^{\infty} \frac{3\left(-
 \lambda e^{-s\hat{\mu}_B}\right)^k}{k^3}
  +4\pi^2 \sum_{k=1}^{\infty}
 \frac{\left(- \lambda e^{-s\hat{\mu}_B}\right)^k}{k} \right]
 \nonumber \\
 &= \frac{2 c}{27\pi^2} \sum_{s=\pm 1} s \left[ 3 \mathrm{Li}_3(-
 \lambda e^{-s\hat{\mu}_B}) -4\pi^2 \log\left(1 +  \lambda
 e^{-s\hat{\mu}_B} \right) \right]\nonumber.
\end{align}
Now, just as in the Stefan-Boltzmann limit (Eq.~\eqref{eq:expsum}), the singularity structure of  $\chi_1^{B,\text{CEM}}$ can be readily  deduced using the analytic properties of the polylogarithms.
The two expressions differ only in the prefactor
$c(T)$, and the fugacity parameter $\lambda(T)$. The latter, being less than unity, shifts the location of the branch cuts away from the imaginary $\hat{\mu}_B$ axis, into the complex $\hat{\mu}_B$ plane.
Thus, the singularities nearest to $\hat{\mu}_B=0$ are located at
$\hat{\mu}_B=\pm \log\lambda \pm i\pi$.
This closely resembles the effect of a nonzero quark mass on the thermal branch points.

\begin{figure}[!tb]
 \centering
 \includegraphics[width=0.99\linewidth]{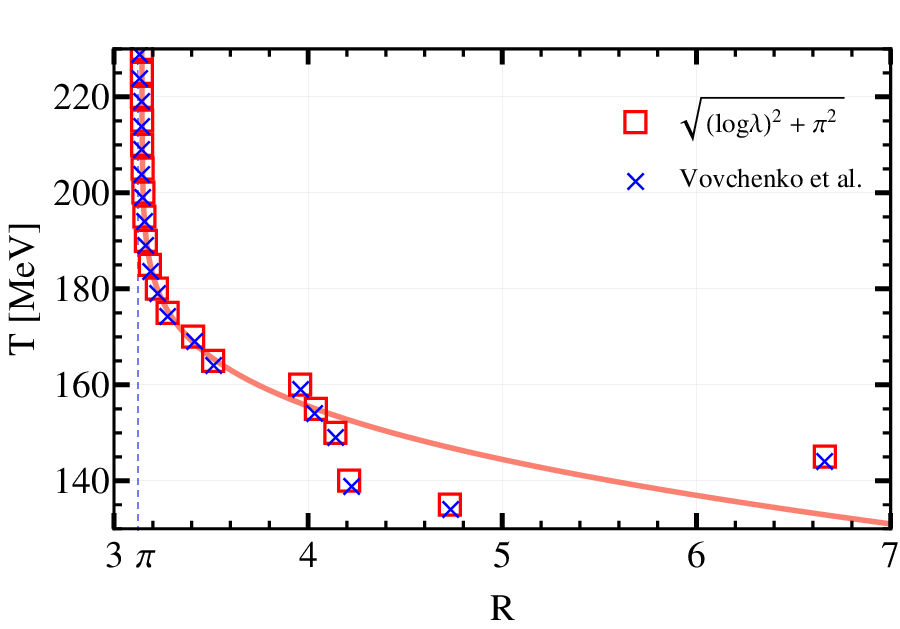}
 \caption{The radius of convergence computed numerically in Ref. \cite{Vovchenko:2017gkg}  and
 the distance of the nearest singularity of the polylogarithm  to the origin (\ref{radius}). The solid
 curve was obtained using the fitted values (\ref{eq:fit-Fourier}) for $b_1$ and $b_2$, as explained in the text.\label{fig:r}}
\end{figure}

The radius of convergence of a Taylor expansion about $\mu_B=0$
is given by the distance to the closest singularity,
\begin{equation}
 \label{radius}
  R=\sqrt{(\log\lambda(T))^2+\pi^2}.
\end{equation}
In Fig.~\ref{fig:r} we show $R(T)$ computed using the Fourier coefficients obtained in LQCD as input. The results obtained with Eqs.~\eqref{radius} and \eqref{eq:par2} for each lattice point
coincide with those obtained in Ref.~\cite{Vovchenko:2017gkg} by summing the Fourier series numerically and using the Mercer-Roberts estimator~\cite{Mercer} of the radius of convergence.
It is thus clear, that the CEM  does not contain information on singularities in the chemical potential plane with $|\text{Im}\,\hat{\mu}|<\pi$. In other words, the CEM exhibits only the
singularities of Eq.~\eqref{eq:expsum2}, which are associated with the periodicity in the imaginary part of the baryon chemical potential.

{Similarly,
 the Fourier series of the RFM
 can be also resummed and expressed by the polylogarithm functions.
Consequently,  neither CEM nor  RFM  can provide any insight into  the location of the  chiral  transition   or on the  existence of the QCD  critical point.  However,  phenomenological models   like CEM can be used
as a useful parametrization of noncritical quantities in QCD.

As we discuss in the following section, criticality is reflected in characteristic properties of high-order Fourier coefficients, which  are not reproduced  by the CEM or RFM.}

\section{Fourier coefficients in a QCD-like effective model}\label{sec:PQM}

In the previous section, we have demonstrated that the Fourier
coefficients $b_k$ cannot be fixed uniquely by knowledge of the first few.
In this section we explicitly examine the effects of critical behavior on the Fourier coefficients within a QCD-like chiral effective model. For this purpose we employ the two-flavor Polyakov-loop quark-meson
model~\cite{Karsch:2010hm,schaefer07:_phase_polyak}, which captures characteristics of QCD at imaginary baryon chemical
potential~\cite{morita11:_role_of_meson_fluct_in}.

\begin{figure}[!t]
 \centering
 \includegraphics[width=\columnwidth]{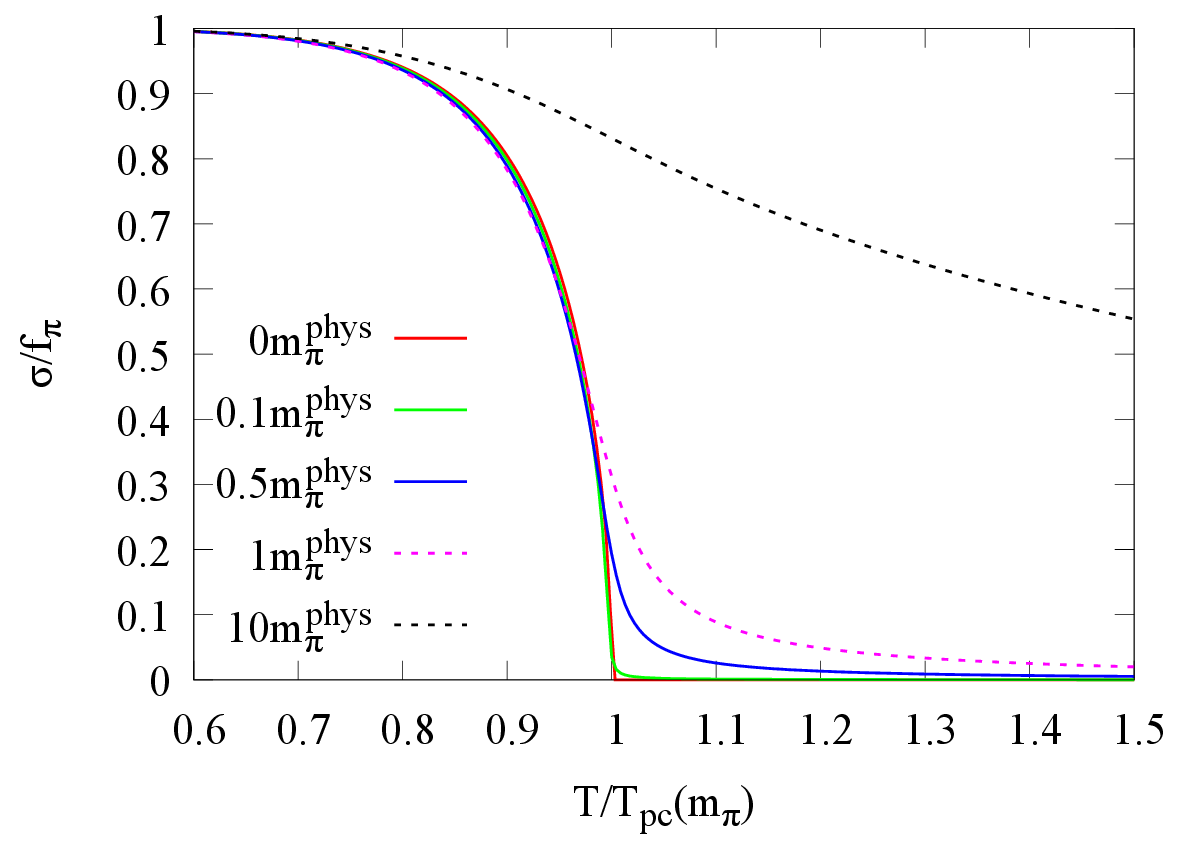}
 \caption{The chiral order parameter in the PQM model for several values of the pion
 mass. The temperature is normalized by the
 (pseudo) critical temperature $T_{pc}(m_\pi)$, for each value of the pion mass. In the chiral limit, $T_{pc}(m_\pi=0)\equiv T_c$.}
 \label{fig:sigma-T}
\end{figure}

\begin{figure*}[!tb]
 \centering
 \includegraphics[width=0.32\textwidth]{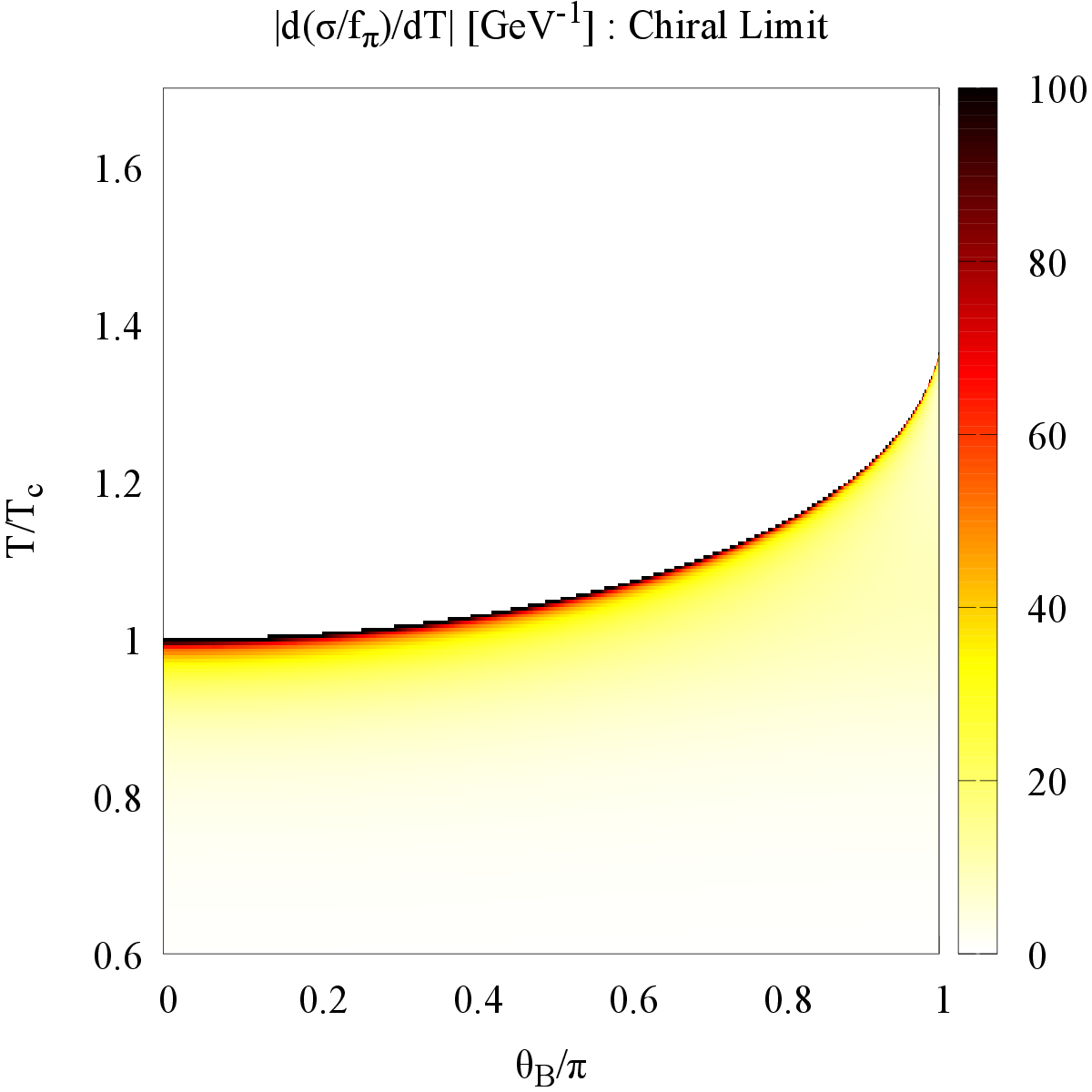}
 \includegraphics[width=0.32\textwidth]{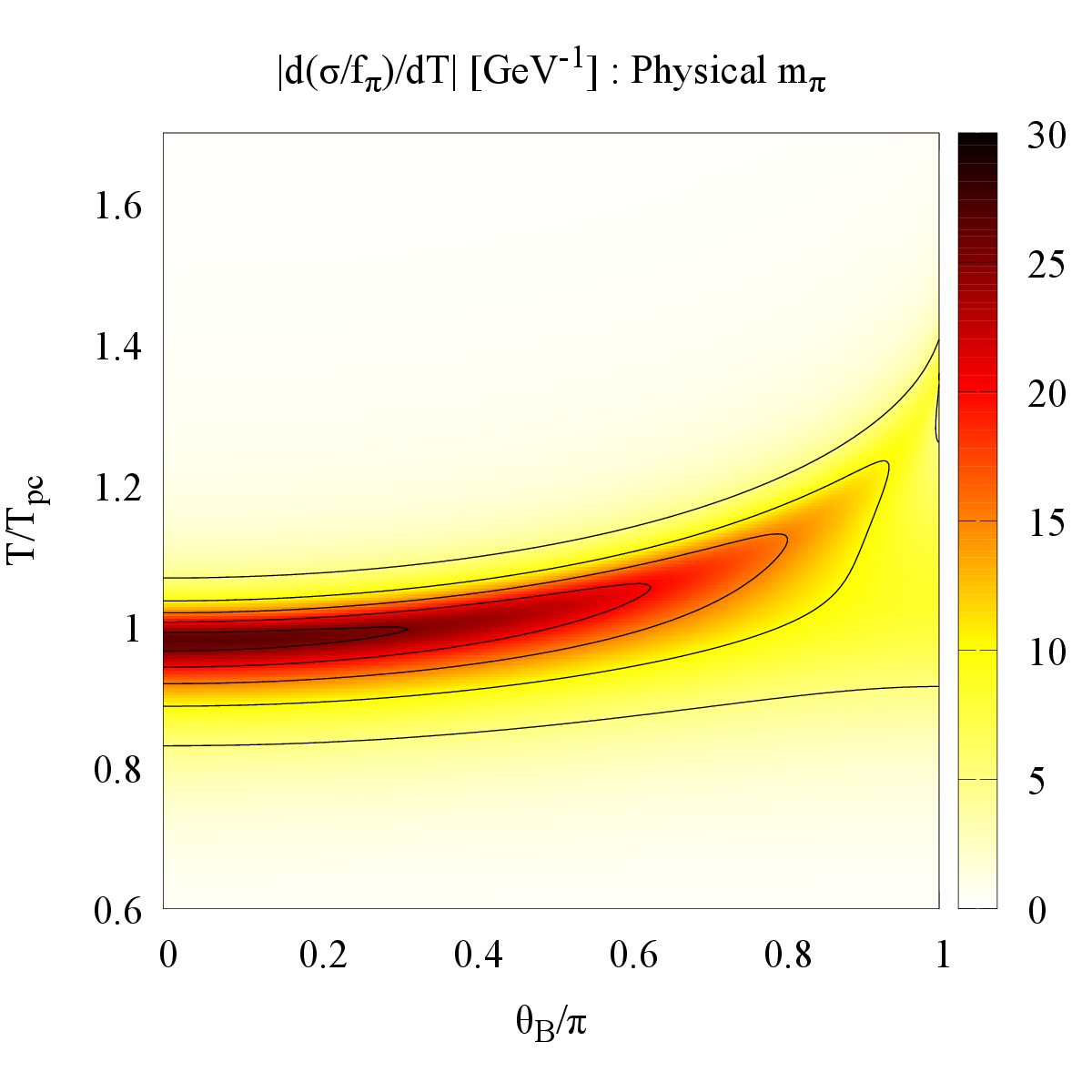}
 \includegraphics[width=0.32\textwidth]{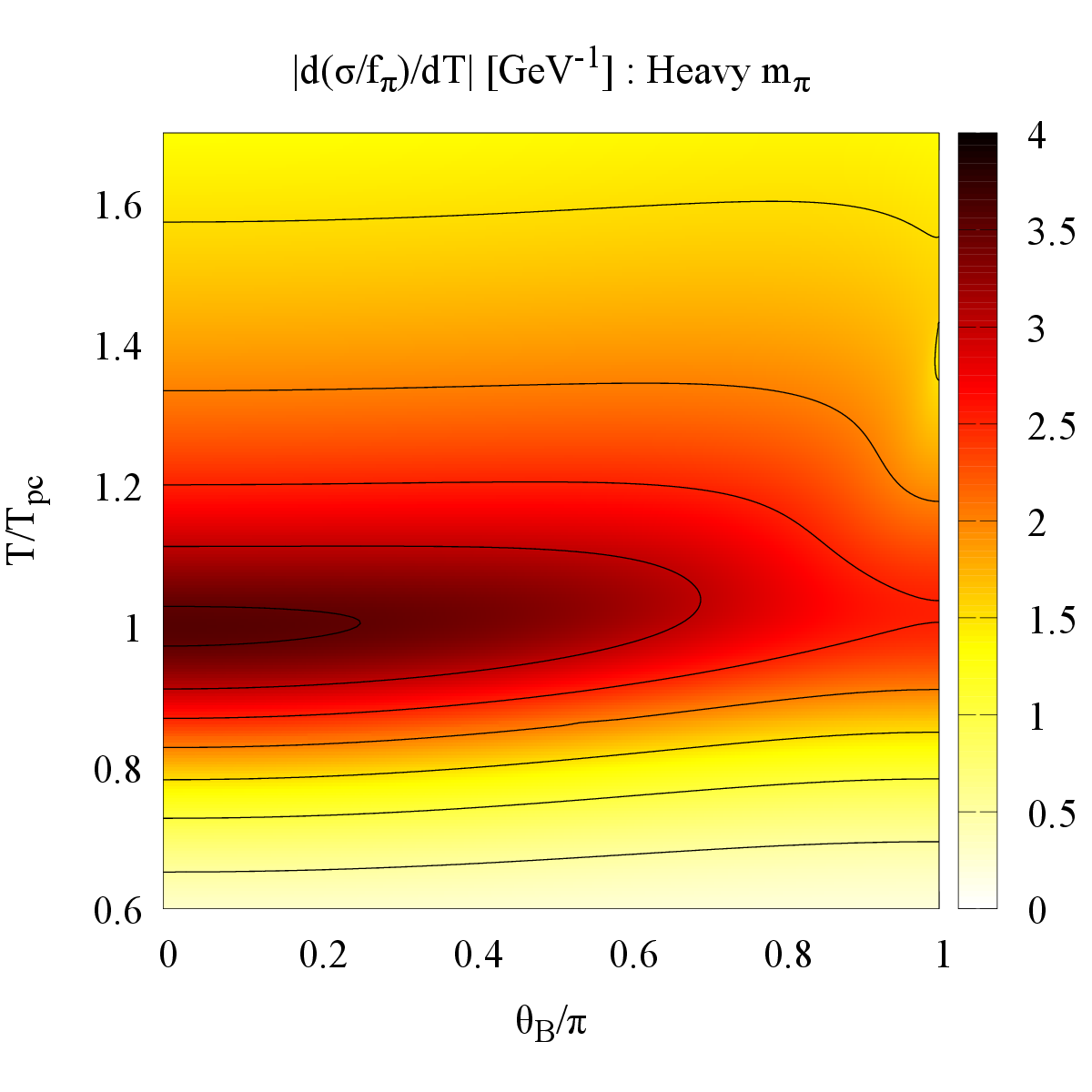}
 \caption{Contour maps for $d(\sigma/f_\pi)/dT$ in the $T-\theta_B$
 plane. The left, middle and right panels correspond to the vanishing, physical and the heavy pion
 mass cases, respectively.}
 \label{fig:thetadep}
\end{figure*}

\subsection{Criticality in the PQM model}

In the PQM model, the thermodynamic potential at nonvanishing temperature $T$ and imaginary quark chemical
potential $\theta_q=\text{Im}\,\mu_q/T$  exhibits  the Roberge-Weiss
periodicity~\cite{roberge86:_gauge_qcd}, a residual of the $Z(N_c)$ symmetry, which is an exact symmetry of QCD in the limit of infinitely heavy quarks. The Polyakov--Nambu--Jona-Lasinio model
\cite{fukushima04:_chiral_polyak,ratti06:_phases_qcd}
is expected to yield compatible results, since the  phase structures, including the Roberge-Weiss transition, are very similar in the two models
~\cite{sakai08:_polyak_nambu_jona_lasin,morita11:_probin_decon_in_chiral_effec}.

In the following we adopt the mean-field approximation, including renormalized quark vacuum fluctuations~\cite{skokov10:_vacuum_fluct_and_therm_of_chiral_model}. The thermodynamic
potential per unit volume $\Omega$ in the PQM model is obtained by extremizing the functional

\begin{equation}
 \Omega(\sigma,\Phi,\Phi^*)= \mathcal{U}(\Phi, \Phi^*)+U(\sigma)+\Omega_{q\bar{q}}(\sigma,\Phi,\Phi^*).
\end{equation}
with respect to the thermal expectation values of the Polyakov loop $\Phi$,
its conjugate $\Phi^*$, and the sigma field $\sigma$.
Here $\mathcal{U}(\Phi,\Phi^*)$ and $U(\sigma)$ are
the Polyakov loop potential and the purely mesonic potential for the $O(4)$
multiplet $(\sigma,\vec{\pi})$, while $\Omega_{q\bar{q}}$ denotes the quark contribution to the thermodynamic potential. We employ the polynomial form for $\mathcal{U}$
\cite{ratti06:_phases_qcd},
\begin{equation}
 \frac{\mathcal{U}(\Phi, \Phi^*)}{T^4} = -\frac{u_2(T)}{2}\Phi^* \Phi -
  \frac{u_3}{6}(\Phi^3 + \Phi^{*3}) + \frac{u_4}{4}(\Phi^* \Phi)^2,\label{eq:poly_pot}
\end{equation}
with
\begin{equation}
 u_2(T) = a_0+a_1\left(\frac{T_0}{T}\right) +
  a_2\left(\frac{T_0}{T}\right)^2+a_3\left(\frac{T_0}{T}\right)^3.
\end{equation}
The parameters in the potential are adjusted so as to reproduce the equation
of state of the pure gluonic matter with $u_3=0.75$, $u_4=7.5$, $a_0=6.75$, $a_1=-1.95$,
$a_2=2.625$, $a_3=-7.44$, and $T_0=270$ MeV. The mesonic potential is given by
\begin{equation}
 U(\sigma) = \frac{\lambda}{4}(\sigma^2-v^2)^2-h\sigma
\end{equation}
where the pion field is suppressed, since we do not consider pion condensation. The explicit chiral symmetry
breaking parameter equals  $h=f_\pi m_\pi^2$.

The quark contribution consists of a vacuum
fluctuation part and a purely thermal part
\begin{align}
 \Omega_{q\bar{q}}(\sigma,\Phi,\Phi^*) &= -\frac{N_c
  N_f}{8\pi^2}m_q^4\ln\left(\frac{m_q}{M}\right)  \\\nonumber
 &-2 N_f T \int\frac{d^3p}{(2\pi)^3} \ln(g^+) + \ln(g^-),
\end{align}
where $M$ is an arbitrary renormalization scale and
\begin{align}
 g^+(\sigma,\Phi,\Phi^*,T,\mu_q) &= 1+3\Phi e^{-(E_q-\mu_q)/T} \\\nonumber
 & +3\Phi^*  e^{-2(E_q-\mu_q)/T}+e^{-3(E_q-\mu_q)/T}
\end{align}
and
\begin{equation}
 g^-(\sigma,\Phi,\Phi^*,T,\mu_q) = g^+(\sigma,\Phi^*,\Phi, T, -\mu_q).
\end{equation}
Here the quark  mass and energy are given by $m_q=g\sigma$ and $E_q=\sqrt{p^2+m^2}$, respectively.

The Polyakov loop variables $\Phi$ and $\Phi^*$ take real values for real
$\mu_q$, such that one can pick $L=|\Phi|$ and $\bar{L}=|\Phi^*|$ as the two independent variables. On the other hand, for imaginary values of the chemical potential they are complex conjugates of each other. Thus, the two independent variables are conveniently chosen as the modulus $L$ and the phase $\phi$, $\Phi=Le^{i\phi}$ and $\Phi^*=Le^{-i\phi}$. The expectation values are then determined by the stationarity condition
\begin{equation}
 \frac{\partial \Omega}{\partial \sigma} = \frac{\partial
  \Omega}{\partial L} = \frac{\partial \Omega}{\partial \phi} = 0,\label{eq:gap}
\end{equation}
and all thermodynamic quantities can be obtained from the thermodynamic potential $\Omega(T,\mu_q)$.

With $N_c=3$ and $N_f=2$, the vacuum parameters are fixed to be $f_\pi=93$ MeV,
$m_\pi^{\text{phys}}=138$ MeV, and $m_\sigma=600$ MeV, while the Yukawa coupling is set to $g=3.35$.

In general, the pseudocritical temperature of a crossover transition is not uniquely determined.
By maximizing the chiral $(\chi_\sigma)$ and Polyakov
loop $(\chi_L)$ susceptibilities, we find, at $\mu_q=0$, the crossover temperatures 231~MeV and 213~MeV for the chiral and deconfinement transitions, respectively. An alternative determination, obtained by maximizing the temperature derivatives of the order parameters, $d\sigma/dT$ and
$dL/dT$, yields $226$~MeV and $223$~MeV  respectively. In the chiral limit the two procedures yield a unique chiral critical temperature.

In the following, we  control the strength of the explicit symmetry breaking  by varying the
pion mass in vacuum from $m_\pi=0$ (chiral limit) to
$m_\pi=10\, m_\pi^{\text{phys}}$, in order to assess the effect of criticality on the Fourier coefficients. Thereby, we keep the same vacuum values for $f_\pi$ and $m_\sigma$ by readjusting
the parameters in $U_\sigma$. The resulting chiral order parameter at
$\mu_B=0$ is shown in Fig.~\ref{fig:sigma-T}. The chiral critical
temperature in the chiral limit is found to be $228.256$ MeV. For a finite,
but small, pion mass $(0.1\, m_\pi^{\text{phys}})$
the pseudocritical temperatures obtained using $\chi_\sigma$ and $d\sigma/dT$ are
229 and 227~MeV, respectively.
Hereafter, for simplicity,  we denote by $T_{pc}$ the pseudocritical temperature corresponding to a maximum of $\chi_\sigma$. However, in the  the heavy pion mass case, the chiral pseudocritical temperature~\footnote{In this particular
    case, $\chi_\sigma$ yields an unusually small transition temperature owing to the sigma-meson mass being much lighter than that of the  pion.}
$T_{pc}=228$~MeV is determined by the maximum of $d\sigma/dT$.

The behavior of order parameters and the phase structure in the
chiral effective models have  been investigated in
Refs.~\cite{sakai09:_deter_qcd,morita11:_probin_decon_in_chiral_effec,morita11:_role_of_meson_fluct_in}.
For the present study  we summarize a few relevant features.
In Fig.~\ref{fig:thetadep}, we show contour maps for the temperature
derivative of the chiral order parameter $d(\sigma/f_\pi)/dT$ in
the plane of imaginary chemical potential vs temperature in the chiral limit and for two nonzero values of the pion mass.

In the left panel of Fig.~\ref{fig:thetadep}, one can unambiguously
identify the chiral critical line
in the chiral limit, which extends to higher temperature at large
$\theta_B$,  and merges with the Roberge-Weiss
transition line at $\theta_B=\pi$ at the critical
temperature $T_c(\theta_B=\pi)=$311~MeV.
This first-order transition line extends from $T=\infty$ down to the
so-called Roberge-Weiss end point,
which exhibits a second-order phase transition at
$(T=T_{\text{RW}}, \theta_B=\pi)$. We find $T_\text{RW}=308.1$ MeV
$=1.35\,T_c$ which is slightly below $T_c(\theta_B=\pi)$.
At $T > T_\text{RW}$, the thermodynamic quantities resemble
qualitatively those of a Stefan-Boltzmann gas.

In the case of a nonvanishing pion mass  (the middle and the right panel in Fig.~\ref{fig:thetadep}),
the rapid change of the order parameter found for a small pion mass is smoothed, such that the region around
the maximum indicates the location of the crossover transition.
Moreover, we find that the curvature of the pseudocritical line is reduced at
small $\theta_B$ as the pion mass is increased.

In LQCD calculations, it is found that the nature of the Roberge-Weiss end point depends on the quark mass
\cite{forcrand:_const_qcd,Bonati:2016pwz}. Thus, for small quark masses it is a triple point (the junction of three first-order phase transitions), while at intermediate values of the quark mass, it is a $Z(2)$ critical point. Finally, for large quark masses, the RW end point is again a triple point. In the model considered here, the RW end point is located at $T_{\text{RW}}=311~\text{MeV} \simeq 1.35\,T_{pc}$ and 327~MeV $\simeq 1.43\, T_{pc}$ for the physical and heavy values of the pion mass, respectively, and is second order in both cases.\footnote{See also  Ref.~\cite{morita11:_probin_decon_in_chiral_effec},
where different forms of the Polyakov loop potential $\mathcal{U}(\Phi,\Phi^*)$ were explored.}

Given the phase  structure in the plane of imaginary chemical potential vs temperature, shown in Fig.~\ref{fig:thetadep},  one expects that the Fourier coefficients $b_k$ at intermediate temperatures, $T_{pc} \leq  T \leq T_{\text{RW}}$, reflect the existence of the transition
line.

\begin{figure}[!t]
 \centering
 \includegraphics[width=\columnwidth]{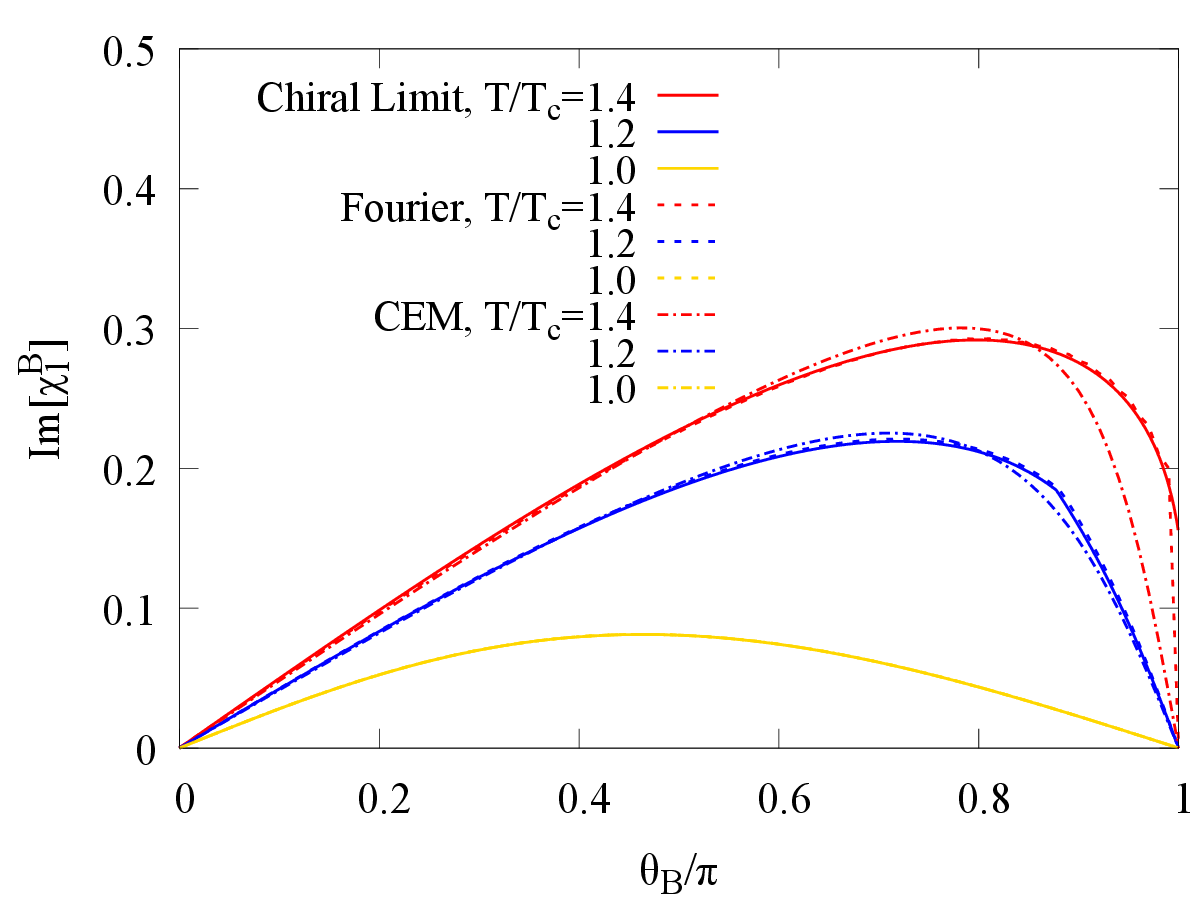}
 \caption{The imaginary part of the baryon density at imaginary chemical
 potential in the chiral limit. Solid lines are obtained by differentiating   the
 pressure with respect to $\hat{\mu}_B$. The dashed lines represent the density reconstructed from the
 Fourier series and the dashed-dotted lines  denote the density reconstructed  from the Fourier coefficients obtained using the CEM ansatz.}
 \label{fig:density}
\end{figure}

\subsection{The Fourier coefficients in the chiral limit\label{sec:chilim}}

In order to identify the influence of criticality on the Fourier coefficients, we
first consider the PQM model formulated in the
chiral limit.
The expansion coefficients $b_k$ are computed by Fourier
transforming the density as a function of the imaginary chemical
potential, as in Eq.~\eqref{eq:bk_Fourier}.
The numerical procedure applied to compute $b_k$  was tested
on the Stefan-Boltzmann gas, and the results are found to be accurate for Fourier coefficients $b_k$ on the order of  $\sim 10^{-9}$ or larger. We note that the applicability of the procedure is limited by the magnitude of the coefficients rather than their order $k$.

In Fig.~\ref{fig:density} we show the normalized baryon density
$\text{Im}\chi_1^B$ in the PQM model calculated in the chiral limit
at imaginary chemical potential and at several temperatures near $T_c$. The solid lines represent the density obtained directly
from the  PQM thermodynamic potential by differentiating with respect to the chemical potential, $\chi_1^B = \partial (p/T^4)/\partial \hat{\mu}_B$.
Also shown are the densities  reconstructed from the
Fourier coefficients obtained in the PQM model using Eq.~\eqref{eq:imagdens}. Finally, we also show the reconstructed densities obtained by using the CEM ansatz \eqref{eq:CEMdef} applied to the PQM model.

At $T=T_c$, the densities coincide,  as is expected from the smooth behavior of the density. However,
at $T/T_c=1.2$ (blue solid curve), the chiral phase boundary, shown in Fig.~\ref{fig:density},  appears as a kink at $\theta_B/\pi \simeq
0.88$. While the kink is reproduced by the density reconstructed from the Fourier coefficients, the CEM ansatz yields a smooth, analytic function. This is expected, since the CEM Fourier coefficients do not capture the power-law behavior of the higher-order coefficients, which is needed to reproduce the singularity.

At $T/T_c=1.4$, the density does not vanish at $\theta_B=\pi$.
This, together with the fact that the density is an odd function of $\theta_B$ that exhibits the Roberge-Weiss periodicity, implies that it must be  discontinuous at this point. The discontinuity is a manifestation of the first-order
Roberge-Weiss transition \cite{roberge86:_gauge_qcd}.

Since $\sin(k\,\theta_B)$ vanishes at $\theta_B=\pi$ for any integer $k$, the density reconstructed from a finite number of terms in the Fourier series \eqref{eq:imagdens} must also vanish at
$\theta_B=\pi$. We  note that the density reconstructed from the  CEM coefficients shows a stronger deviation near $\theta_B=\pi$. Moreover, even the closed form  expression
\eqref{eq:expsum2} for the resummed CEM Fourier series, does not reproduce the discontinuity in the density for $\lambda<1$, i.e., $T<\infty$. Thus, we expect significant differences between high-order $b_k$ and $b_k^{\text{CEM}}$ at temperatures beyond $T_\text{RW}$.
In the following,  we explore the structure of the Fourier coefficients $b_k$ in more detail.

\begin{figure}[!t]
 \centering
 \includegraphics[width=\columnwidth]{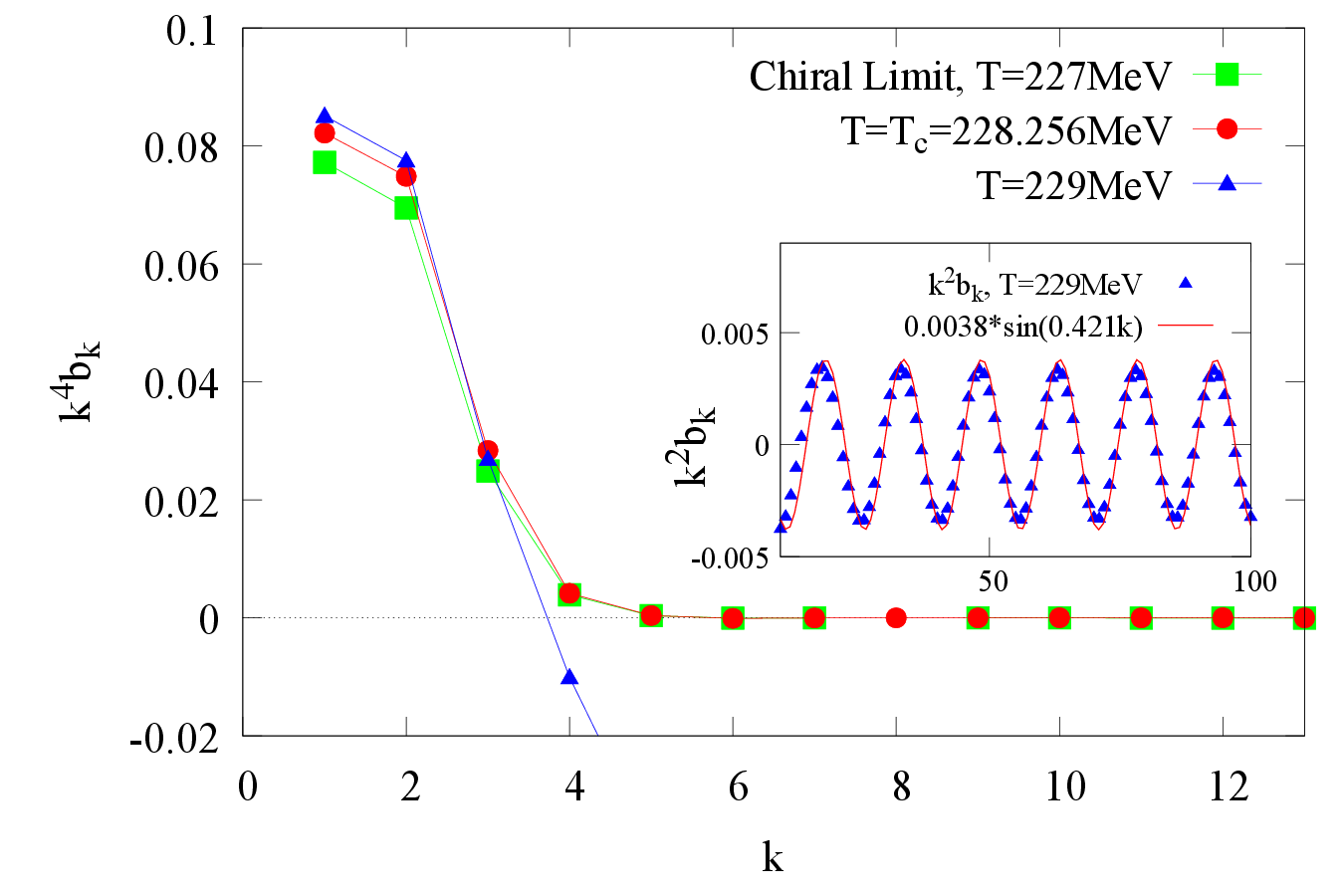}
 \caption{Fourier coefficients $k^4b_k$ in the chiral limit at temperatures near $T_c$. The  inset shows $k^2 b_k$ at a temperature slightly above $T_c$, $T=229$ MeV. The solid line represents a fit with $A\sin(k\theta_c)$, as discussed in the text. }
 \label{fig:bk_Tc_0m}
\end{figure}

At temperatures significantly below $T_c$, the PQM model exhibits statistical
confinement. Consequently there is an oscillatory dependence of the thermodynamic potential on the
imaginary baryon chemical potential, with
$\Delta\Omega(T,\theta)\sim \cos(3\theta_q) = \cos\theta_B$
\cite{morita11:_probin_decon_in_chiral_effec}, as in
the hadron resonance gas. Therefore the density is, to a good approximation,  that of a
classical Boltzmann gas, $\text{Im}\chi_1^B \sim \sin\theta_B$ with small corrections from quantum statistics and interactions.
This, in turn, implies that the higher-order Fourier coefficients are strongly suppressed, as in the CEM. Our results confirm this expectation up to $T/T_c \leq 0.9$ for Fourier coefficients of order $k<9$. Owing to the numerical limitations mentioned above, coefficients of higher order, i.e., $k\gtrsim 10$, cannot be computed reliably in this temperature range, within the present scheme.

Given the results on the signature of the Roberge-Weiss transition in the Fourier series, one may expect that the chiral critical line at real values of the baryon chemical potential~\cite{skokov11:_mappin,friman12:_phase_trans_at_finit_densit,stephanov06:_qcd_critic_point_and_compl} could also contribute significantly to the high-order Fourier coefficients. However, contributions to the Fourier coefficients from singularities located at real $\mu_B$ are exponentially suppressed~\cite{Fourier_general}, i.e.,  $\Delta b_k \sim e^{-k\,|\text{Re}\,{\hat{\mu}_B}|}$. This leads to an exponential suppression also of the contribution from the thermal branch points discussed in Sec.~\ref{sec:analytic} for nonzero fermion masses, with $\Delta b_k \sim e^{-k\, m}$.

From general scaling considerations one finds that at $T=T_c$ and at large $k$,
mean-field criticality leads to an asymptotic dependence of  $b_k\sim 1/k^4$  \cite{Fourier_general}.
However, since in this case the integration contour, except for the point $\theta_B=0$, lies in the chirally broken phase, a major contribution to the Fourier coefficients is due to massive fermion degrees of freedom. Consequently, the initial $k$-dependence of the Fourier coefficients is exponential, while the contribution of the critical point $\theta_B$ is fairly small, implying that the power law is visible only at very large $k$.

Figure \ref{fig:bk_Tc_0m} shows $b_k$ around $T=T_c=228.256$ MeV. Here
we have multiplied  $b_k$ by  $k^4$ in order to highlight the large $k$ behavior of the Fourier coefficients. One observes a rapid suppression of $b_k$ with increasing order
below and at $T_c$, in line with a substantial contribution of massive degrees of freedom to the baryon density.
Thus, at temperatures $T\leq T_c$, the density is saturated by the first few Fourier coefficients. Consequently, the CEM ansatz is able to reproduce the baryon density at $T=T_c$, as shown in Fig.~\ref{fig:density}, although the model does not yield the correct asymptotic behavior of the Fourier coefficients. Moreover, owing to the dominant contribution of massive degrees of freedom, it is extremely difficult, if not impossible, to reliably extract the asymptotic $1/k^4$ behavior of the coefficients in numerical calculations.

At a temperature slightly above $T_c$, we find that $b_k$
exhibits oscillations, with a $1/k^2$ dependence superimposed, as shown in the inset in Fig.~\ref{fig:bk_Tc_0m} for $T=229$ MeV.
This characteristic dependence on $k$ is expected at temperatures between the critical temperature at $\mu_B=0$, $T_c$, and the Roberge-Weiss temperature $T_\text{RW}$. In this temperature range, the critical point is located on the imaginary $\mu$ axis (see Fig.~\ref{fig:complexplane}). It follows that the Fourier coefficients take the asymptotic form
$A\sin(k\theta_c)/k^2$ \cite{Fourier_general}, where $\theta_c=\text{Im}\,\mu_c/T$ is the critical value of the baryon chemical potential.
Indeed, a fit of the oscillatory dependence on $k$ shown in Fig.~\ref{fig:bk_Tc_0m} yields $\theta_c=0.421$, in perfect agreement with the location of the critical point at $T=229$ MeV.

\begin{figure}[!t]
 \centering
 \includegraphics[width=\columnwidth]{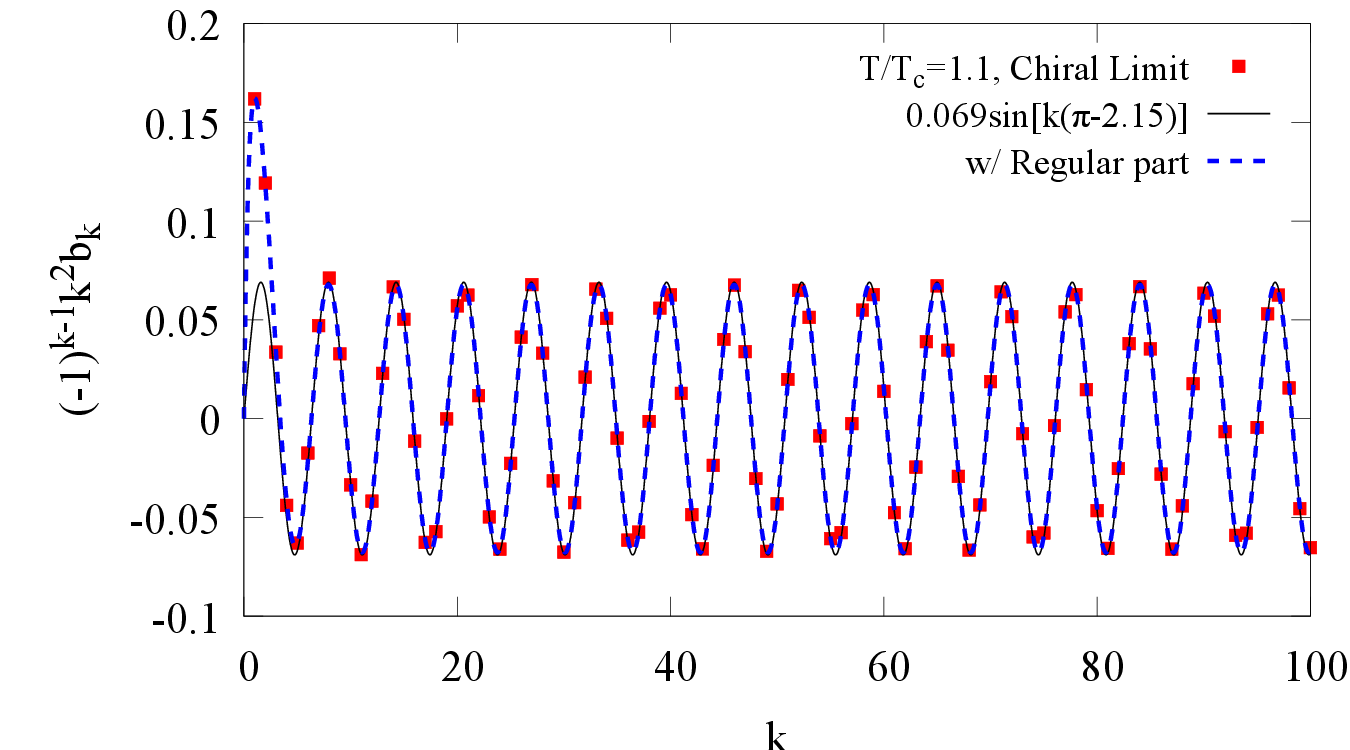}\\
 \includegraphics[width=\columnwidth]{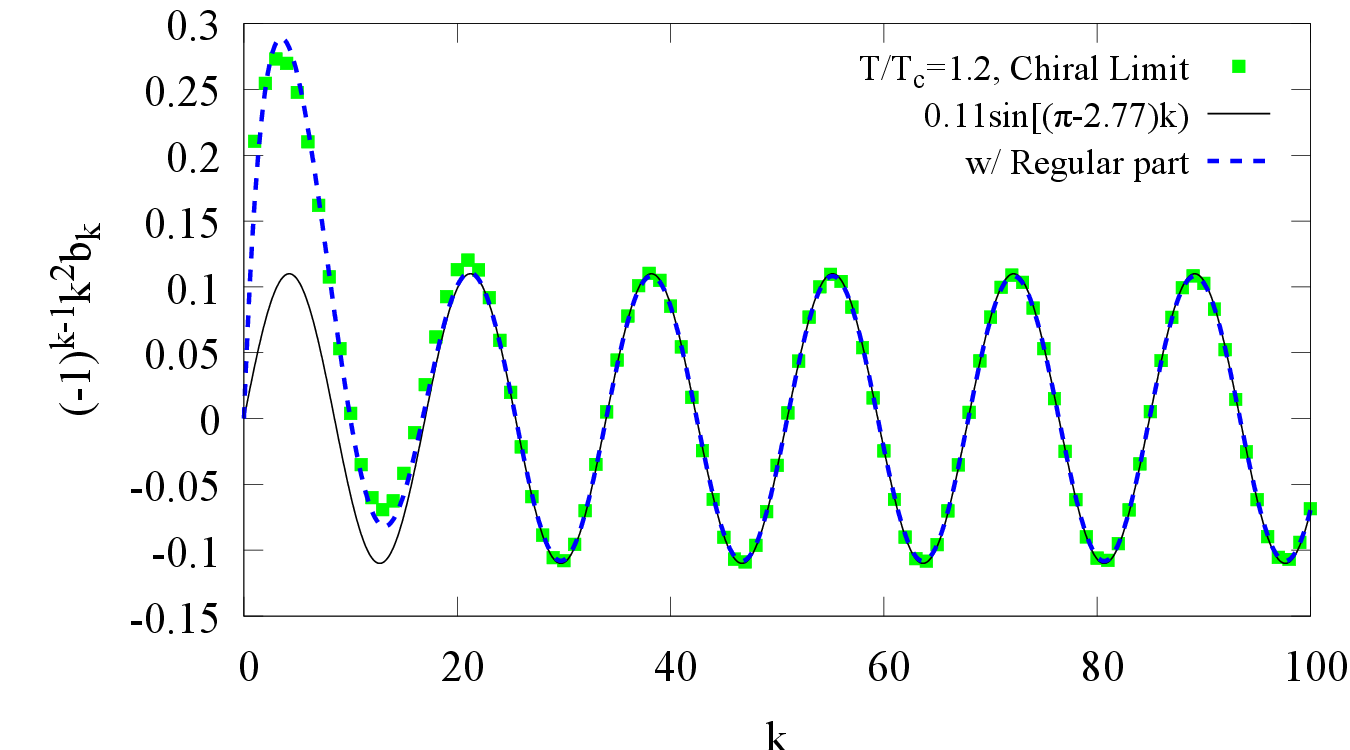}\\
 \includegraphics[width=\columnwidth]{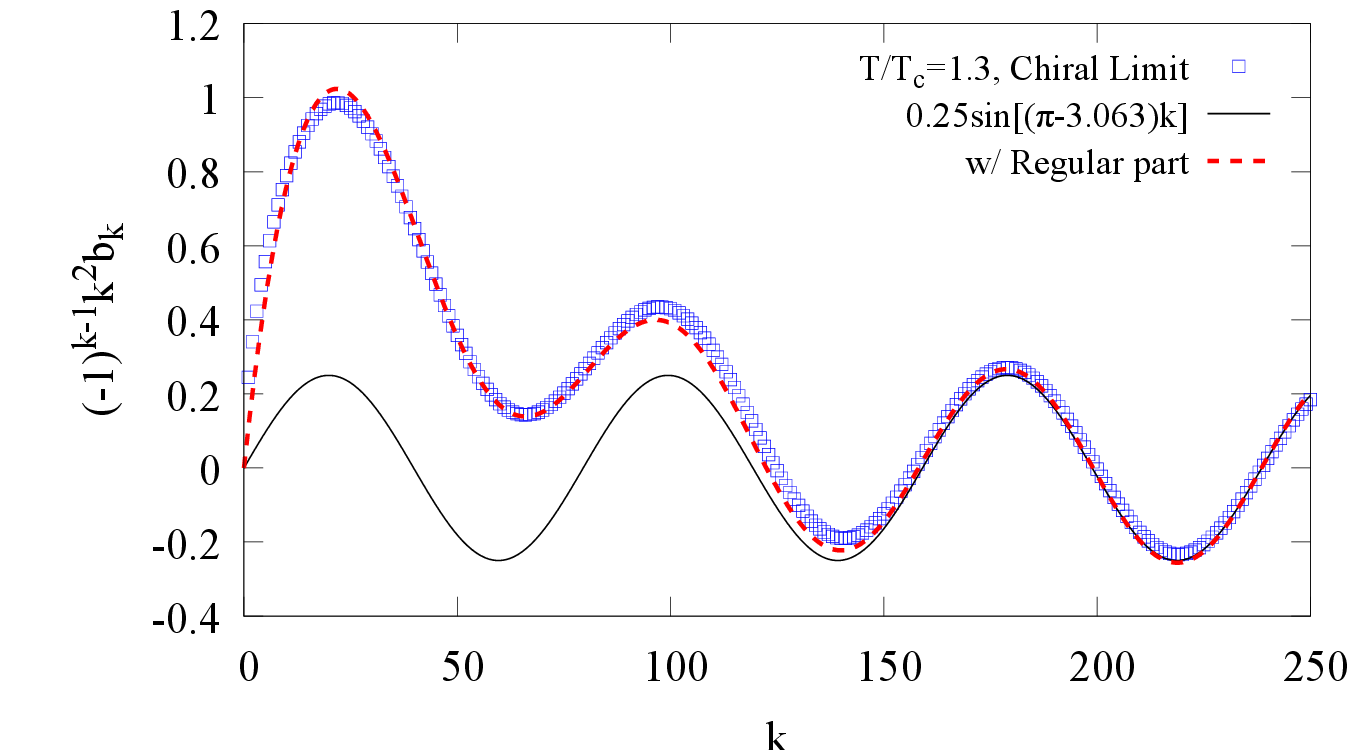}
 \caption{Fourier coefficients of the PQM model in the chiral limit
 at $T/T_c=1.1$ (top), 1.2 (middle), and 1.3 (bottom). The solid lines show
 a fit with $A\sin[(\pi-\theta_c)k]$.}
 \label{fig:chilim_aboveTc}
\end{figure}

The expectation that the oscillation frequency depends on the location of
the singularity is confirmed by the behavior of $b_k$ at higher temperatures. In Fig.~\ref{fig:chilim_aboveTc} we show the Fourier coefficients at $T/T_c=1.1$, 1.2, and 1.3.
Here we have multiplied $b_k$ by  $(-1)^{k-1}$. The additional phase, changes the frequency of the
oscillation from $\theta_c$ to $\pi-\theta_c$, and thus yields oscillations that are more easily discernible.
This is because it is difficult to identify a frequency larger than $\pi/2$ on the discrete $k$ grid.
The lines show fits with the functional form
$A\sin[k(\pi-\theta_c)]$. Also shown in Fig.~\ref{fig:chilim_aboveTc}
are fits where a contribution of the regular part of the
baryon density, outside the critical region, is included:
$ C\,K_2(a\,k)/k+A\sin[k(\pi-\theta_c)]$. Here $K_2(x)$ denotes the
modified Bessel function of the second kind.
The  values obtained for $\theta_c=2.15$, 2.77, and 3.063
are in agreement with the location of the chiral critical point for the corresponding temperature.
For instance, $\theta_c=2.77$ at $T/T_c=1.2$ yields
$\theta_c/\pi = 0.88$, which is the position of the kink in the density in Fig.~\ref{fig:density}.

The increase of $\theta_c$ with temperature is a consequence of the shape of the phase
boundary (see Fig.~\ref{fig:thetadep}), which shifts to larger values of $\theta$ with increasing temperature.
We conclude that the high-order Fourier coefficients carry information on the chiral phase transition at imaginary values of the baryon chemical potential.

\subsection{Fourier coefficients at nonzero pion masses\label{sect:nonzero_pion_mass}}

\begin{figure}[!t]
 \centering
 \includegraphics[width=\columnwidth]{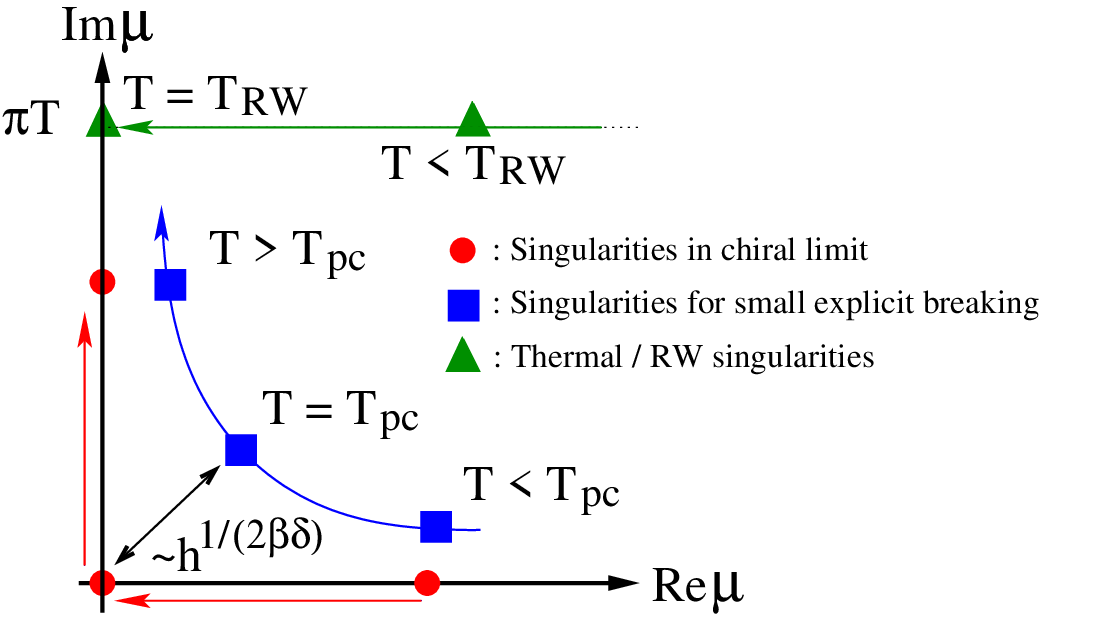}
 \caption{Schematic view of the singularities in the complex chemical
 potential plane (first quadrant only). The arrows indicate the
 direction each singularity moves when the temperature is increased.}
 \label{fig:complexplane}
\end{figure}

\begin{figure}[!tb]
 \centering
 \includegraphics[width=\columnwidth]{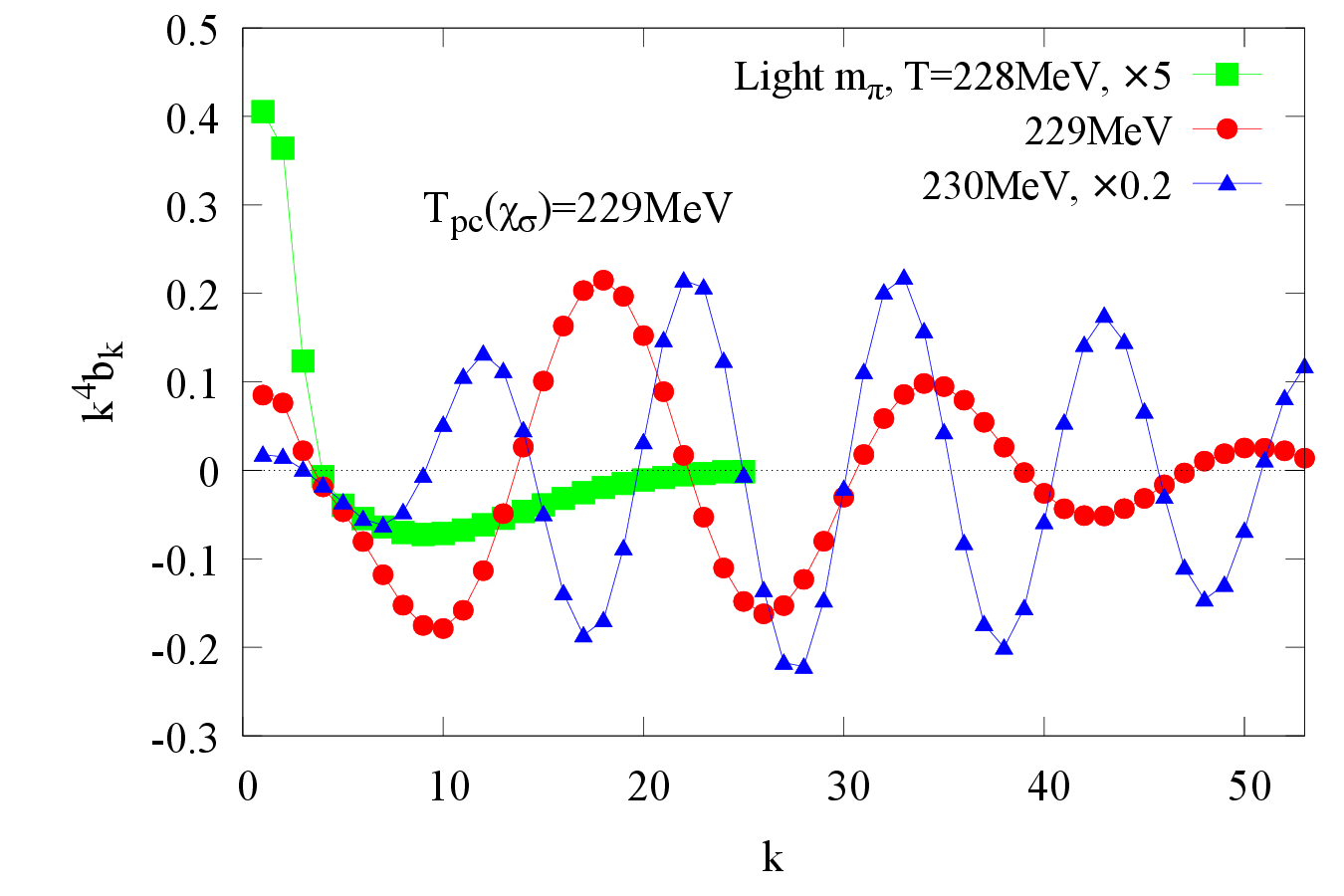}\\
 \includegraphics[width=\columnwidth]{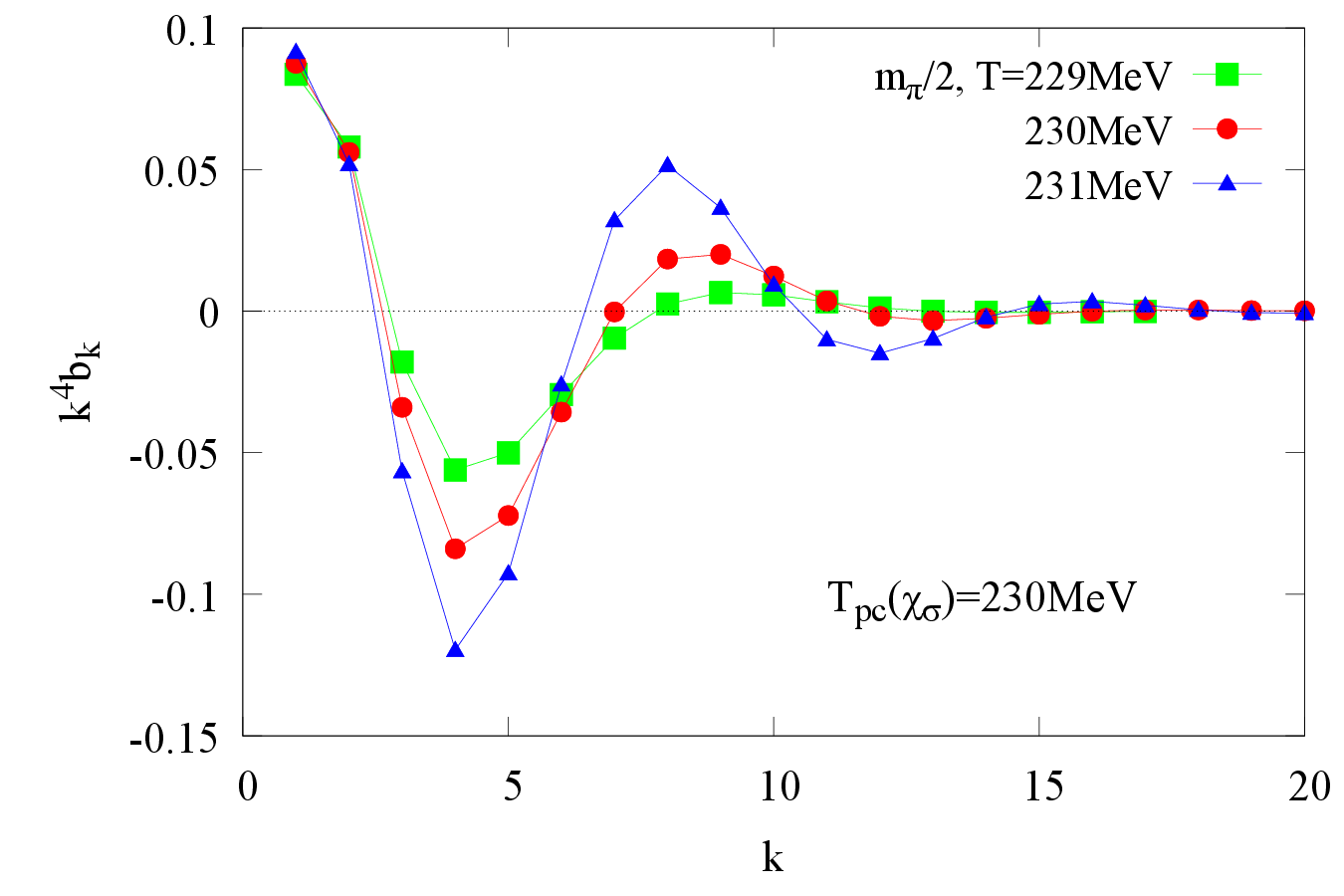}
 \caption{The Fourier coefficients $k^4 b_k$ at $m_\pi=0.1\,m_\pi^{\text{physical}}$ (upper panel) and at
 $0.5\,m_\pi^{\text{phys}}$ (lower panel) for temperatures near $T_{pc}$. In the lower
 panel, the results at $T=228$ and 229~MeV have been multiplied by the factors 5  and 0.2, respectively.}
 \label{fig:bk_Tc_smallmpi}
\end{figure}

\begin{figure}[!t]
 \centering
 \includegraphics[width=\columnwidth]{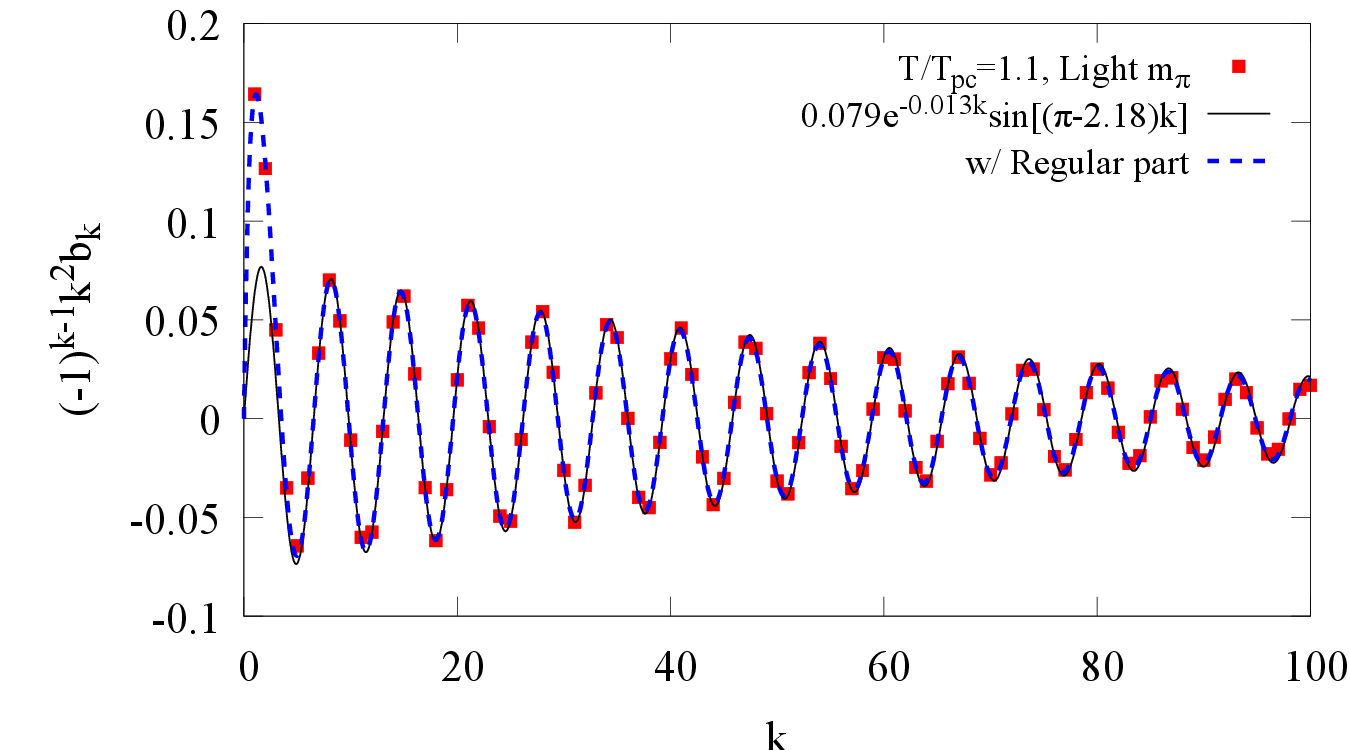}\\
 \includegraphics[width=\columnwidth]{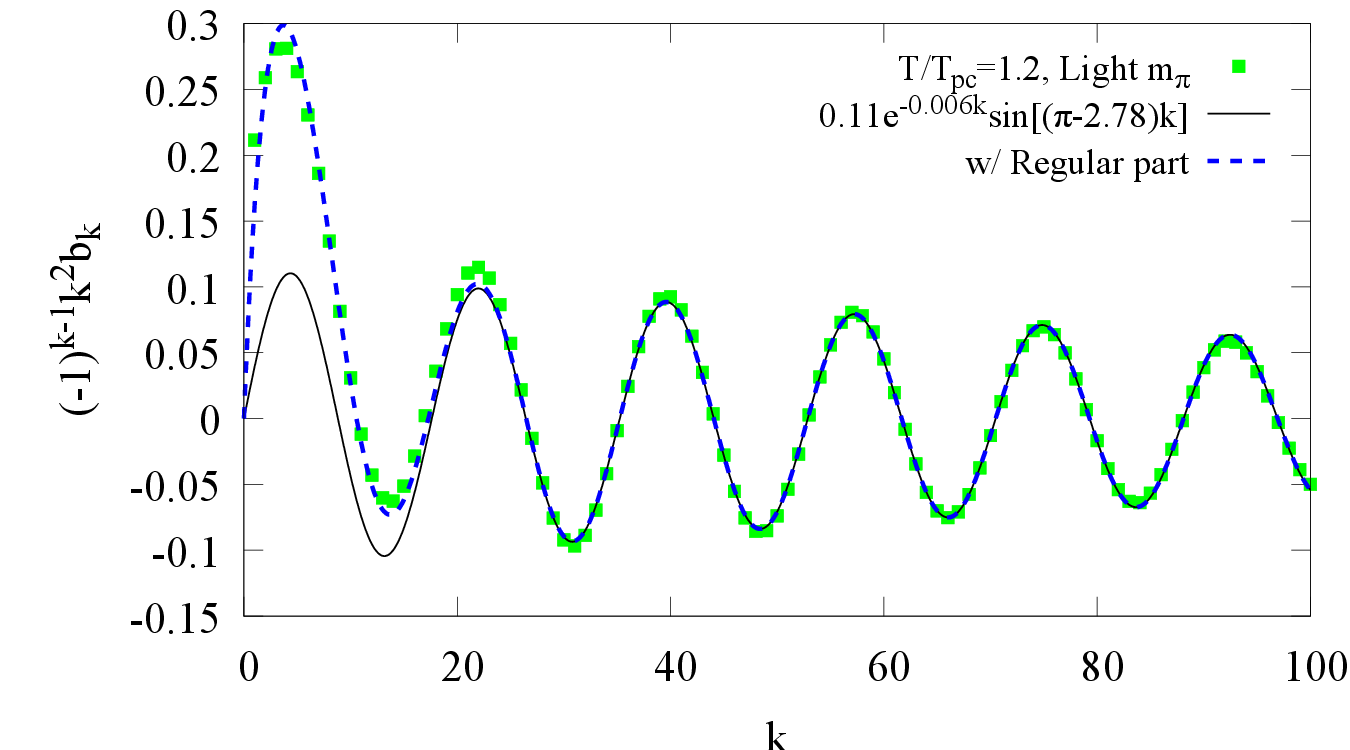}\\
 \includegraphics[width=\columnwidth]{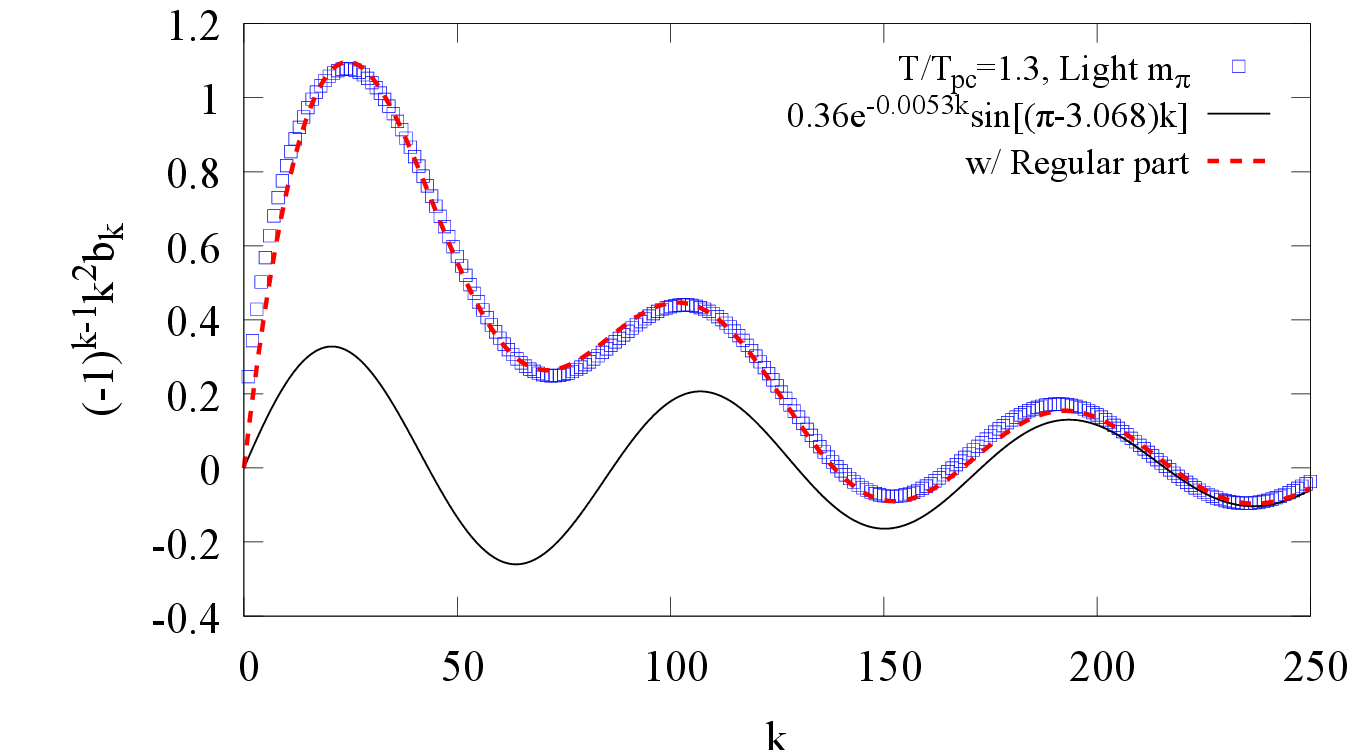}
 \caption{Same as Fig.~\ref{fig:chilim_aboveTc}, but for the light pion mass.}
 \label{fig:lightpi_aboveTc}
\end{figure}

For nonzero pion masses, the chiral symmetry
is explicitly broken, implying that the chiral transition is of the crossover type and that the analytic properties of the baryon
density in the complex chemical potential plane are modified.
Specifically, there is a shift of the chiral critical point from the real or imaginary axis into the complex chemical potential plane~\cite{stephanov06:_qcd_critic_point_and_compl,friman12:_phase_trans_at_finit_densit}, as illustrated in
Fig.~\ref{fig:complexplane}. Owing to the real part of the chemical potential at the singularity, the high-order Fourier coefficients exhibit exponential damping in addition to the power-law scaling and oscillatory behavior found in the chiral limit for temperatures above $T_{pc}$~\cite{Fourier_general}.

\begin{figure*}[!tb]
 \centering
 \includegraphics[width=0.32\textwidth]{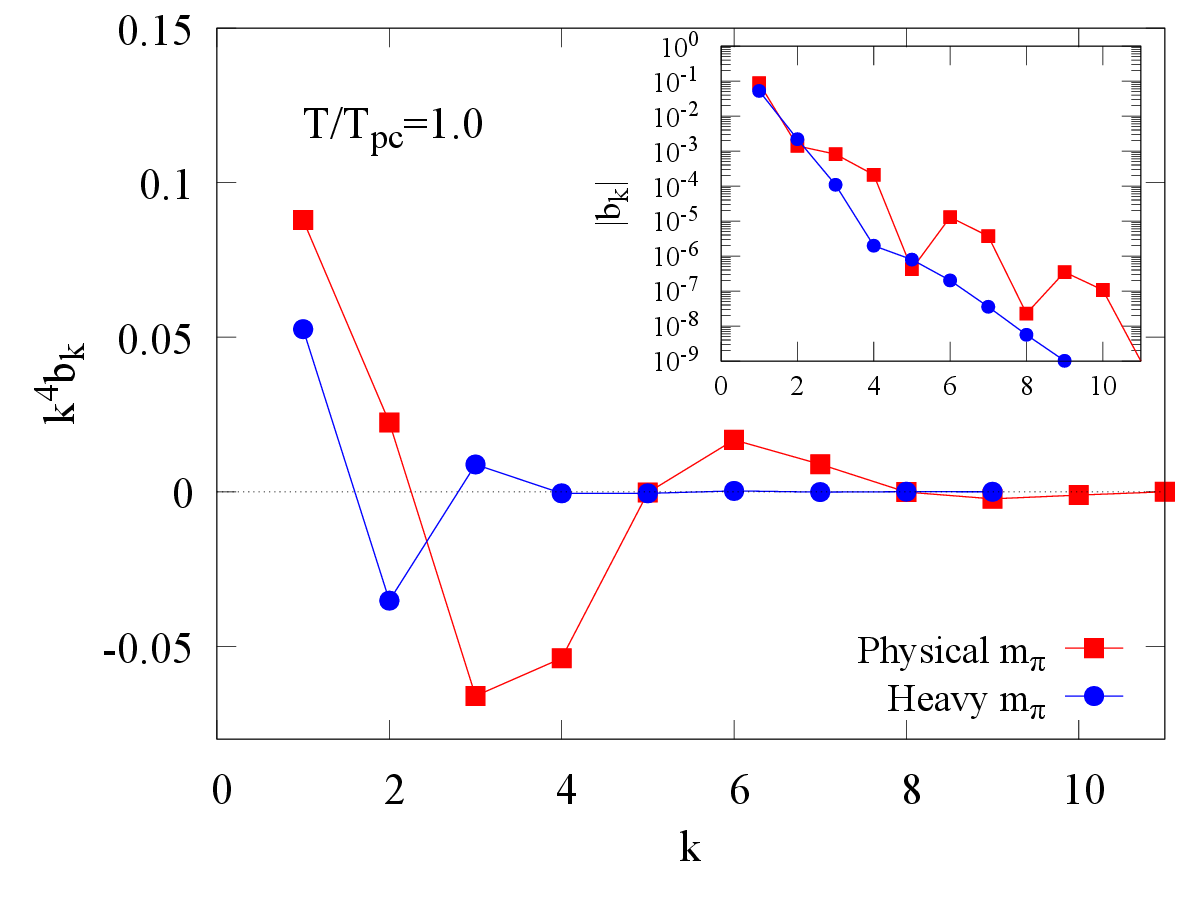}
 \includegraphics[width=0.32\textwidth]{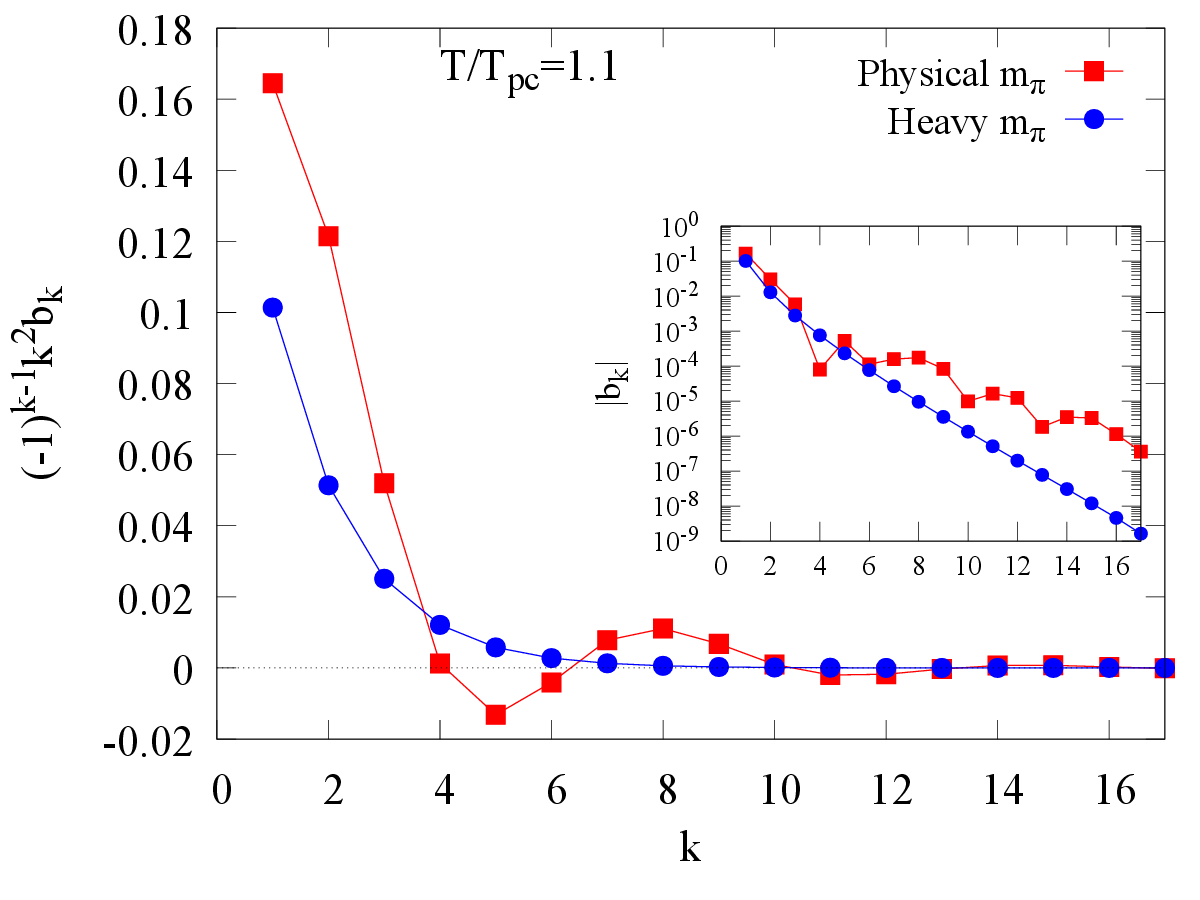}
 \includegraphics[width=0.32\textwidth]{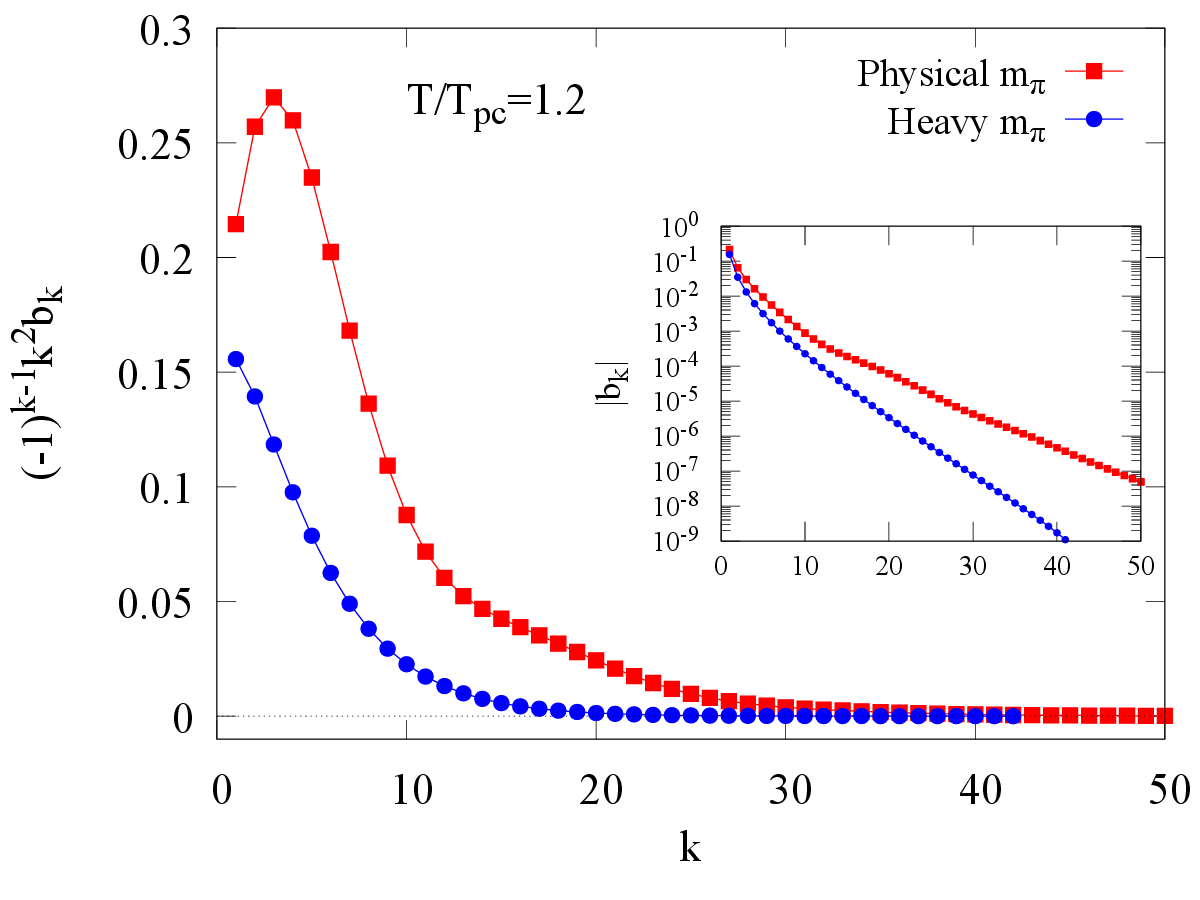}
 \caption{The Fourier coefficients $b_k$ for physical and heavy pion
 masses in the PQM model.  From left to right, the
 results are  shown for $T/T_{pc}=1.0$, 1.1, and 1.2, respectively.
 The insets show the magnitude of $|b_k|$ on a logarithmic scale.
 }
 \label{fig:bk_tc}
\end{figure*}

The resulting Fourier coefficients for temperatures near the crossover temperature $T_{pc}$ are shown in Fig.~\ref{fig:bk_Tc_smallmpi}. The upper panel shows the
results for a small pion mass,
$m_\pi=0.1\,m_\pi^{\text{phys}}$. At $T=T_{pc}$, we find an oscillatory
behavior of $k^4 b_k$,
accompanied with a slow damping of the amplitude.
A comparison with temperatures
slightly  above and below $T_{pc}$, shows that the oscillation frequency
increases with temperature, while the damping of the amplitude is reduced with temperature.
This observation is in qualitative agreement with the temperature dependence of the location  of the singularity in the complex chemical potential plane, shown schematically in Fig.~\ref{fig:complexplane}.

We note that, given the limited range in $k$ available $(k\lesssim 50)$, one cannot distinguish between a  fit with a function proportional to $e^{-ak}/k^2$ with large $a$
from one proportional to  $e^{-ak}/k^4$ and small $a$.
Nevertheless, these results clearly demonstrate the sensitivity of the Fourier coefficients to small changes in temperature and pion mass and thus to chiral criticality.

In the lower panel of Fig.~\ref{fig:bk_Tc_smallmpi}, we show
the Fourier coefficients for a somewhat heavier  pion mass, equal to $0.5\,m_\pi^{\text{phys}}$. Here the temperature dependence of the oscillation frequency is reduced. This is expected, since
the relevant energy scale, the pion
mass, is now large  compared to the few MeV change in the temperature. Nevertheless, the qualitative behavior of the oscillations as a function of temperature is still consistent with the expected motion of the singularity in the complex chemical potential plane. Thus, the frequency increases and the damping of the amplitude weakens with increasing temperature.

Unlike the critical temperature in the chiral limit, the pseudocritical temperature $T_{pc}$ is not clearly distinguished by a sudden change in the properties of the Fourier components $b_k$. This is a rather natural consequence of  the shift of the
singularity into the complex chemical potential plane, indicated in Fig.~\ref{fig:complexplane}. As the temperature is increased and passes  $T_{pc}$, the  motion of the singularity in the complex plane leads to a slow increase of the oscillation frequency and a smooth reduction of the exponential damping of the amplitude.

As in Fig.~\ref{fig:chilim_aboveTc}, we show in  Fig.~\ref{fig:lightpi_aboveTc}
the Fourier coefficients  multiplied by the factor $(-1)^{k-1}k^2$ at
$T/T_{pc}=1.1$, 1.2, and 1.3.
In order to fit the large $k$ dependence of the coefficients $b_k$,  we introduce an exponential damping factor of the form $e^{-a\,k}$, as well as a regular part.
The temperature dependence of the oscillation frequency and the damping
factor is again consistent with the expectations deduced from
Fig.~\ref{fig:complexplane}.

A comparison of the  results for small pion masses with those obtained in the chiral limit, shows that the higher-order Fourier coefficients are exponentially damped by the explicit breaking of chiral symmetry. Therefore, it is a quantitative question whether chiral criticality can be identified in the Fourier coefficients also for larger pion masses. In order to address this issue, we present the Fourier coefficients obtained in the PQM model for the physical and for a heavy pion mass in
Fig.~\ref{fig:bk_tc}.

As illustrated  in
Fig.~\ref{fig:sigma-T}, in the mean-field approximation the PQM  model  exhibits a rather rapid chiral crossover transition for a pion mass equal to the physical one, while LQCD yields a smoother transition at the physical point
\cite{borsanyi10:_in_t_qcd,bazavov12:_chiral_and_decon_aspec_of_qcd_trans}. Therefore, a mean-field calculation with a larger pion mass, somewhere between the physical and the heavy pion mass employed here, may be more relevant for a comparison with lattice results.

In  the left panel of Fig.~\ref{fig:bk_tc} we show  $k^4\, b_k$ for physical (squares)
and heavy (circles) pion masses at $T=T_{pc}$. For a physical pion mass,
the behavior is qualitatively similar to that shown in the lower panel of Fig.~\ref{fig:bk_Tc_smallmpi}.

We note that the sign structure of the Fourier coefficients differs from the alternating signs of the LQCD results (see Fig.~\ref{fig:coeffs} and Ref.~\cite{Vovchenko:2017xad}).  As indicated above, the oscillation frequency depends on the location of the contributing singularities in the complex chemical potential plane, in particular on their imaginary parts, as well as on the strength of the individual contributions. Thus, the sign structure of the Fourier components is determined by an interplay between the chiral singularity and, e.g., the thermal branch point. Consequently, the staggered sign structure seen in the LQCD results at $T=T_{pc}$ may be due to noncritical physics, not captured by the PQM model.

The oscillations lead to deviations from a pure exponential damping of the magnitude of the Fourier coefficients, as shown in the inset of the left panel of Fig.~\ref{fig:bk_tc}, while for the heavy pion mass, the oscillations are almost completely washed out by the strong damping.

In the middle and center panels of Fig.~\ref{fig:bk_tc} we show
$(-1)^{k-1}k^2b_k$ at $T/T_{pc}=1.1$ and 1.2. The additional phase factor compared to the left panel, removes the dominant oscillation to a large extent and hence yields clearer plots. At $1.1\,T_{pc}$, almost all points are at positive values, with a few exceptions for the physical pion mass. Moreover, at $1.2\,T_{pc}$ the sign changes are completely removed by the phase factor. This implies that the staggered sign structure  of the LQCD results on the first four Fourier coefficients~\cite{Vovchenko:2017xad} is reproduced at temperatures somewhat above the pseudocritical temperature $T_{pc}$.

As shown in the insets in the middle and right panels, we indeed find an exponential damping of the magnitude of $b_k$ for the heavy pion mass for $k$ larger than 3 or 4.  At $T=1.2\,T_{pc}$, a small amplitude, low-frequency residual oscillation,  left after removing the dominant one with the phase factor $(-1)^{k-1}$,  would be difficult to  recognize, owing to the fast exponential damping. Nevertheless, the shoulder around $k \sim 10-20$ for the physical pion mass (see inset) may be the signature of such an oscillation.

From the above discussion, it is clear that the location of the singularity associated with the chiral transition in the
complex chemical potential plane is reflected in characteristic $k$-dependencies of the Fourier coefficients $b_k$. However, the presence of other singularities, which interfere with the chiral one, and the exponential damping of the Fourier coefficients for physical pion masses make the analysis less clear cut.

\subsection{The effect of the Roberge-Weiss transition on the Fourier coefficients}

\begin{figure}[!tb]
 \centering
 \includegraphics[width=\columnwidth]{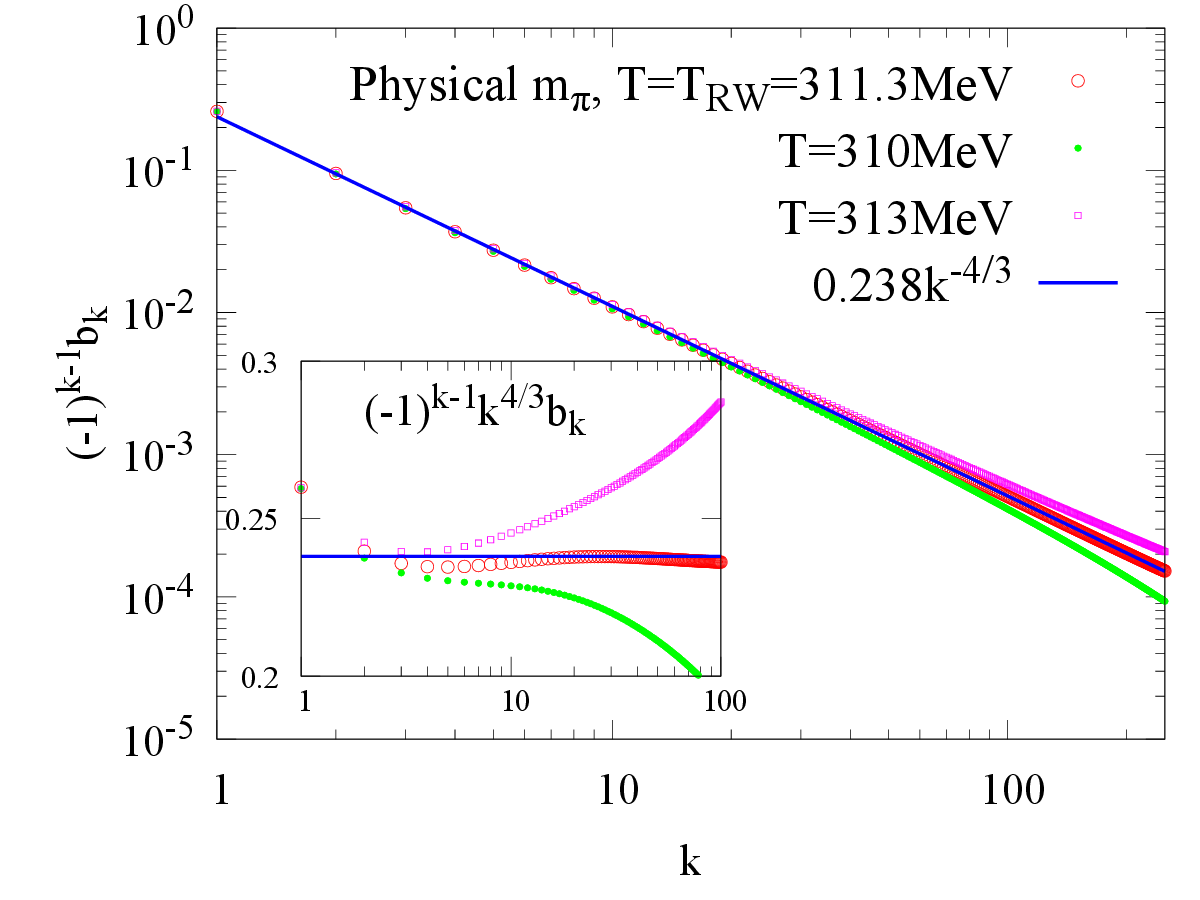}
 \caption{Fourier coefficients near $T=T_\text{RW}$ for the physical pion
 mass. In the inset the coefficients multiplied by $k^{4/3}$ are shown.}
 \label{fig:bk_RW}
\end{figure}

Irrespective of the value of the pion mass, the Roberge-Weiss transition appears at  $T\geq T_{\text{RW}}$. It is a first-order transition, which ends at $T=T_{\text{RW}}$. Assuming that the RW end point is a second-order critical point, one finds that the leading
contribution to the high-order Fourier coefficients at $T=T_{\text{RW}}$ is of the form
$b_k \sim (-1)^{k-1}k^{-(1+1/\delta)}$ \cite{Fourier_general},  where $\delta$ is the critical exponent associated with the external field strength in the $Z(2)$ universality class.
In Fig.~\ref{fig:bk_RW} we show the Fourier coefficients at
$T=T_\text{RW}$ for the physical value of the pion mass.
We find that the coefficients $b_k$ indeed follow a power-law decay with the exponent $4/3$, consistent with $\delta=3$ in the mean-field approximation. In order to illustrate the sensitivity, we also
display $b_k$ at $T$ slightly below and slightly above $T_{\text{RW}}$.
The inset shows that the
deviation from the $k^{-4/3}$ behavior at the slightly different
temperatures is substantial for $k > 10$.
We also note that, in the chiral limit, we find a deviation from the
$k^{-4/3}$ scaling. We attribute this to the
contribution from the chiral phase boundary, which is located close to the RW end point.

\begin{figure}[!tb]
 \centering
 \includegraphics[width=\columnwidth]{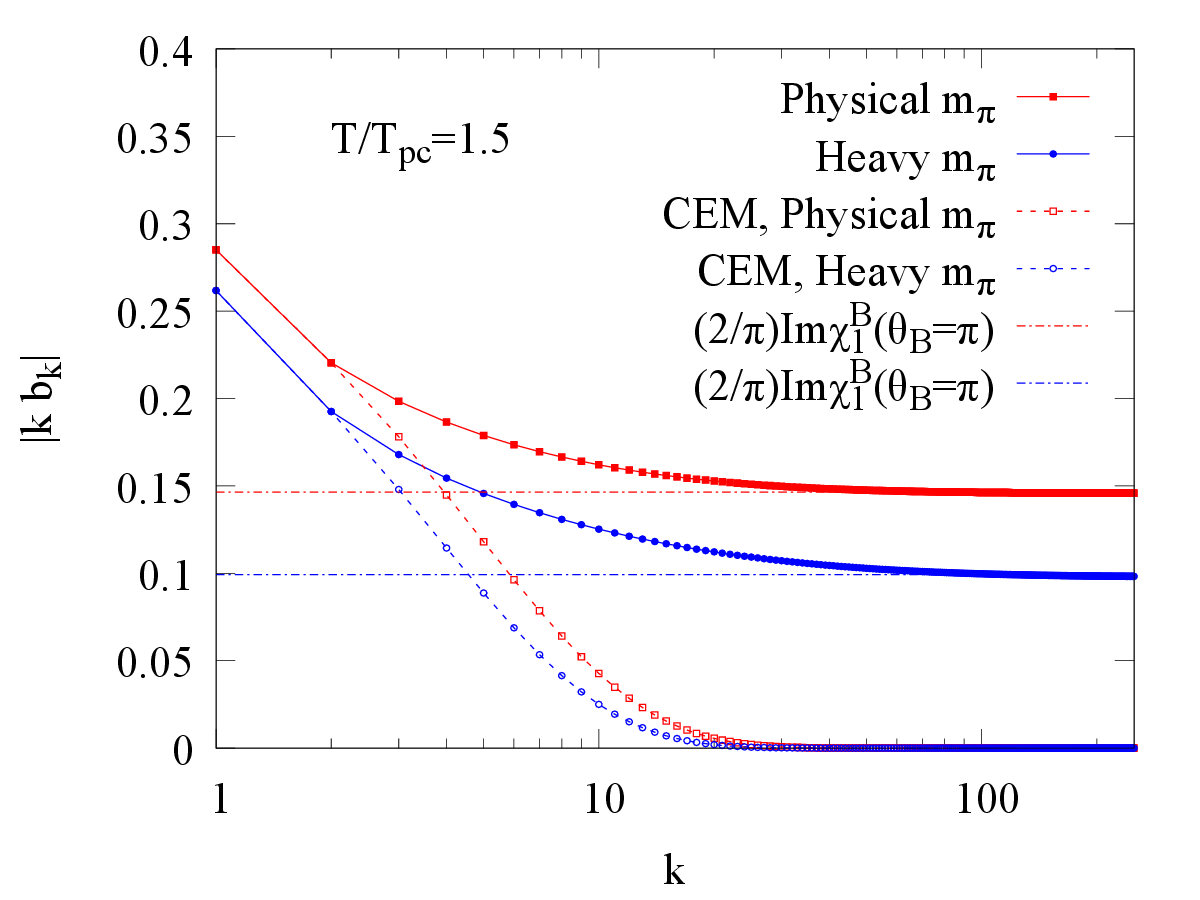}
 \caption{The Fourier coefficients $|b_k|$ for a temperature above $T_{\text{RW}}$
 calculated in  the PQM model and in CEM. The dashed-dotted horizontal lines indicate the expected
 asymptotic value, which is determined by the discontinuity in the density.}
 \label{fig:bk_highT}
\end{figure}

At temperatures above $T_\text{RW}$, the density is discontinuous
at $\theta_B=\pi$, implying that the Fourier coefficients take the asymptotic form~\cite{Fourier_general}  $b_k \sim (-1)^{k-1}/k$.
This is confirmed by Fig.~\ref{fig:bk_highT}, where we plot
$|k\, b_k|$ for $T/T_{pc}=1.5$. The asymptotic value of $k\, b_k$  is given by $(2/\pi)\text{Im}\chi_1^B(T,i\pi)$, and thus directly  connected with the discontinuity in the density at the Roberge-Weiss transition \cite{Fourier_general}.

For comparison we also show the Fourier coefficients obtained using the CEM scheme in Fig.~\ref{fig:bk_highT}. Clearly, the CEM ansatz does not reproduce the asymptotic behavior of the coefficients related with the RW transition, as anticipated in the discussion of Fig.~\ref{fig:density}, in Sec.~\ref{sec:chilim}.  Thus, the CEM ansatz cannot account for the Roberge-Weiss transition, irrespective of the value of the pion mass.

Although these results are obtained in a particular model, the qualitative features are of more general validity, since they are directly linked to the Roberge-Weiss transition.

\subsection{Temperature dependence of low-order coefficients}

\begin{figure*}[!t]
 \centering
 \includegraphics[width=\columnwidth]{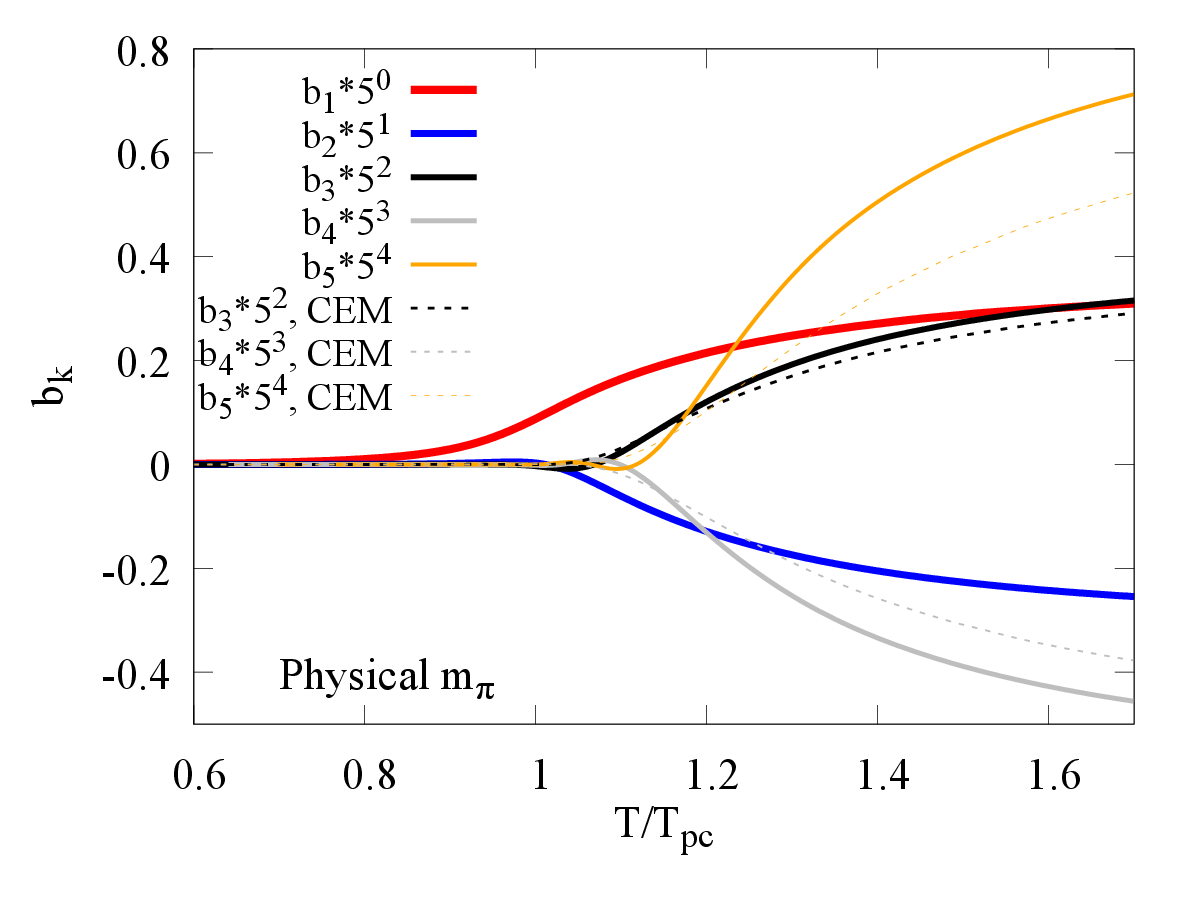}
 \includegraphics[width=\columnwidth]{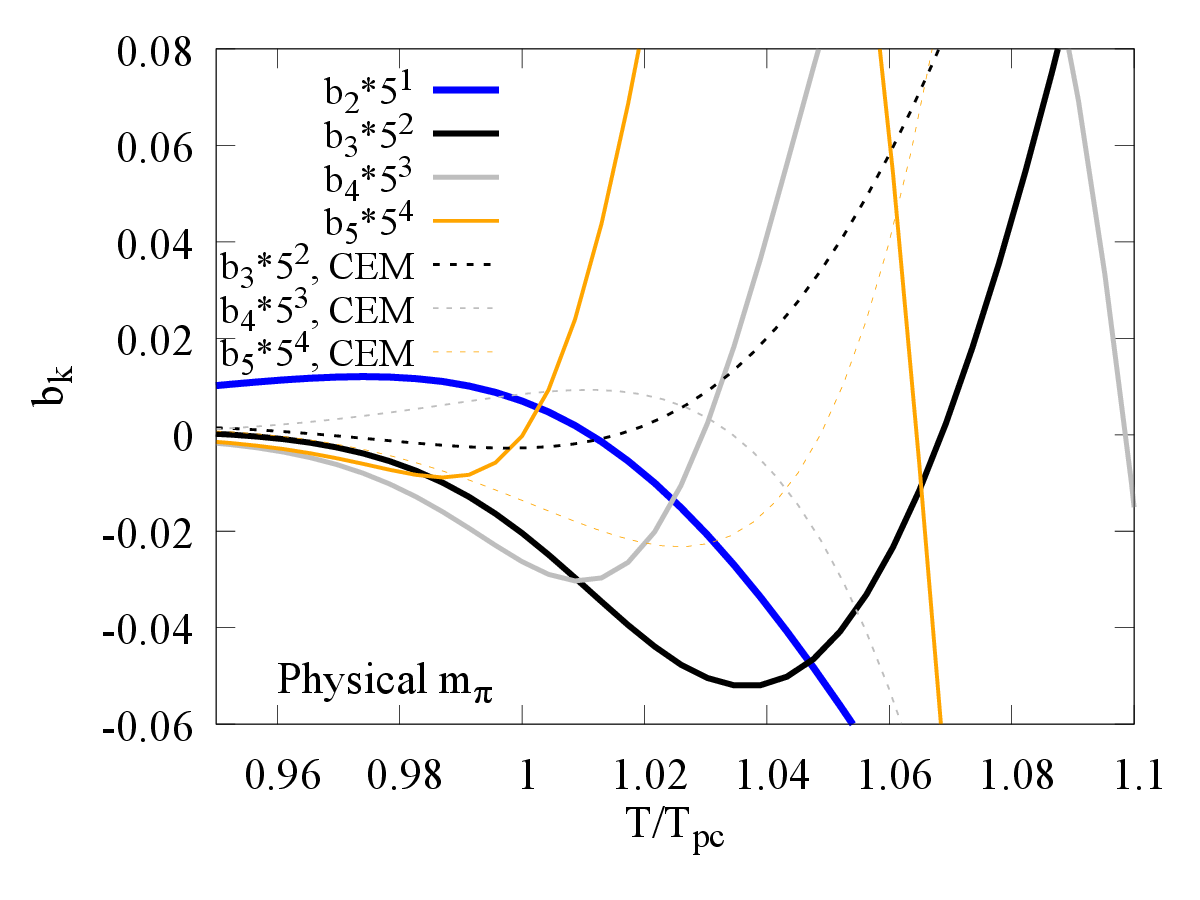}

 \includegraphics[width=\columnwidth]{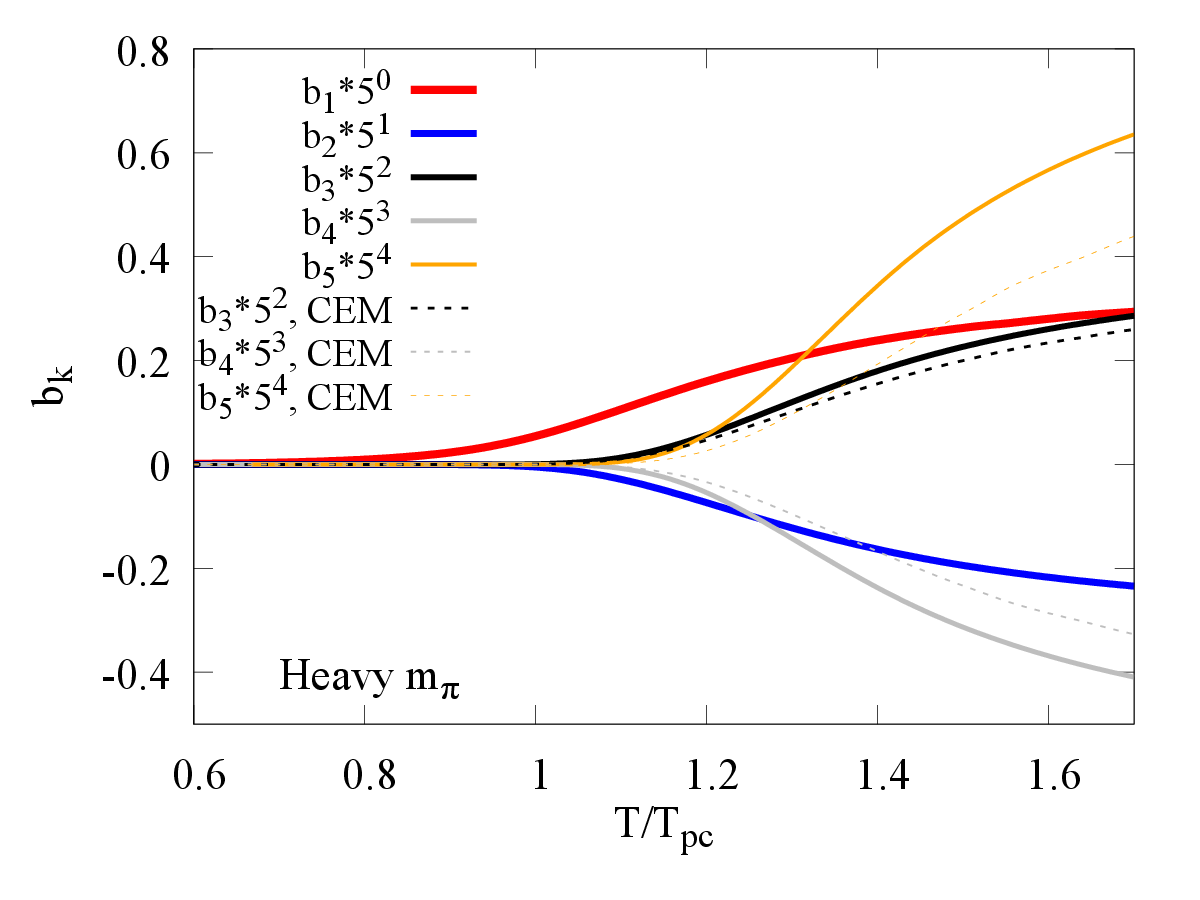}
 \includegraphics[width=\columnwidth]{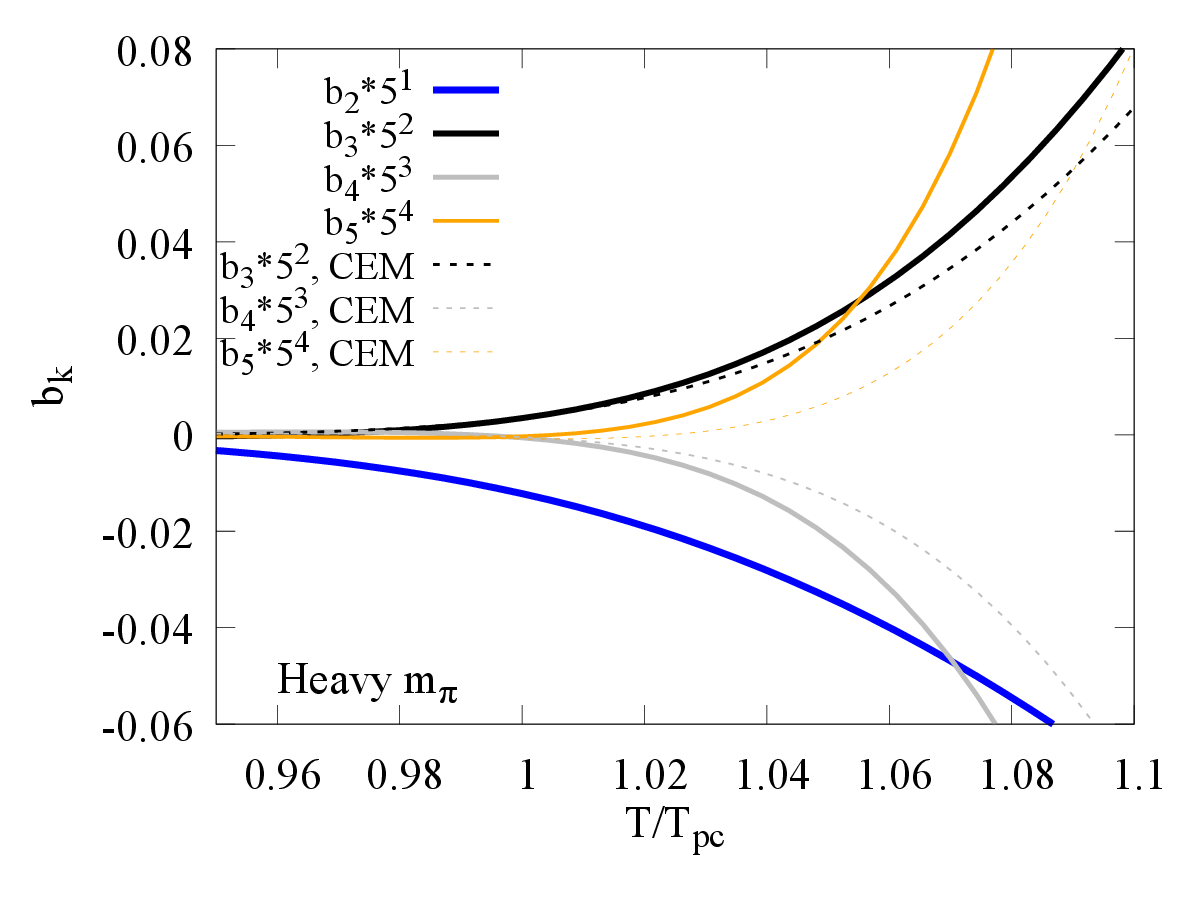}
 \caption{The Fourier coefficients $b_k$ in the PQM model as functions of
 temperature. Solid lines show the PQM model results, while the dashed lines are constructed using the CEM ansatz. The upper
 and lower figures  show  the physical and heavy pion
 mass cases, respectively. In order to improve readability, each $b_k$ is  multiplied with a factor
 $5^{k-1}$.}
 \label{fig:bk-T}
\end{figure*}

A crossover or a true phase transition may also be
signaled by the temperature dependence of the Fourier coefficients.
In Fig.~\ref{fig:bk-T} we show the  $b_k$, up to $k=5$,  as functions
of temperature for physical and heavy pion masses.
For comparison we also show the coefficients obtained using the CEM ansatz.

For the physical pion mass, the chiral crossover transition is reflected
already in the lowest-order coefficients through the oscillations  of $b_k$
for temperatures near $T_{pc}$, as shown in the upper right panel.
On the other hand, for a heavy pion mass this
oscillatory behavior  is suppressed, owing to the stronger smoothening. The coefficients obtained using  the CEM ansatz do not exhibit any oscillatory behavior and show
substantial deviations from the PQM results with increasing temperature.

These results
indicate that already at order $k\geq 2$, the  Fourier expansion coefficients may
exhibit a nontrivial temperature dependence in the crossover region.
We note that the Fourier coefficients obtained in LQCD do not seem to oscillate as functions of temperature~\cite{Vovchenko:2017xad}. However, as discussed in Sec.~\ref{sect:nonzero_pion_mass}, the interplay between critical and noncritical physics, and consequently the sign structure of the Fourier coefficients, depends on the location and relative strength of the singularities, which may be different in the model and in QCD. Consequently, we expect that oscillations of the Fourier coefficients in the crossover region will appear also in LQCD calculations, when the location of the chiral singularity approaches the origin of the complex $\mu$-plane, i.e., for pion masses smaller than the physical value (cf.~ Fig.~\ref{fig:complexplane}).

\subsection{Reconstructing  susceptibilities  from $b_k$}

The characteristic behavior of the Fourier coefficients $b_k$
is reflected also in the baryon number fluctuations.
Using Eq.~\eqref{eq:cumulants}, the cumulants at $\mu_B=0$ can be obtained by using the sum
\begin{equation}
 \chi_{2n}^B = \sum_{k=1}^{k_\text{max}}k^{2n-1}b_k.\label{eq:chin_sum}
\end{equation}
Formally the sum should be extended to infinite order but, provided $b_k$
decreases sufficiently fast with $k$, the series  converges and the summation may be truncated at some moderate value $k_{\text{max}}$.

We find that for $n\leq 8$ the baryon number cumulants $\chi_{2n}^B$  computed using Eq.~\eqref{eq:chin_sum}  reproduce those obtained  directly
by taking derivatives of the pressure  $\chi_{n}^B = \partial^n (p/T^4)/\partial \hat{\mu}_B^n$ for temperatures up to
$T/T_{pc} \leq 1.1-1.3$. For higher temperatures it is numerically difficult to
obtain reliable results for the  coefficients $b_k$  at sufficiently high order to reach agreement.  We note  that there is an intrinsic problem in calculating higher-order cumulants  using the Fourier expansion at temperatures above $T_{\text{RW}}$, owing to the  slow
decrease of the $b_k$  with $k$, which is related to the discontinuity in density at the RW transition.

\begin{figure*}[!t]
 \centering
 \includegraphics[width=0.32\textwidth]{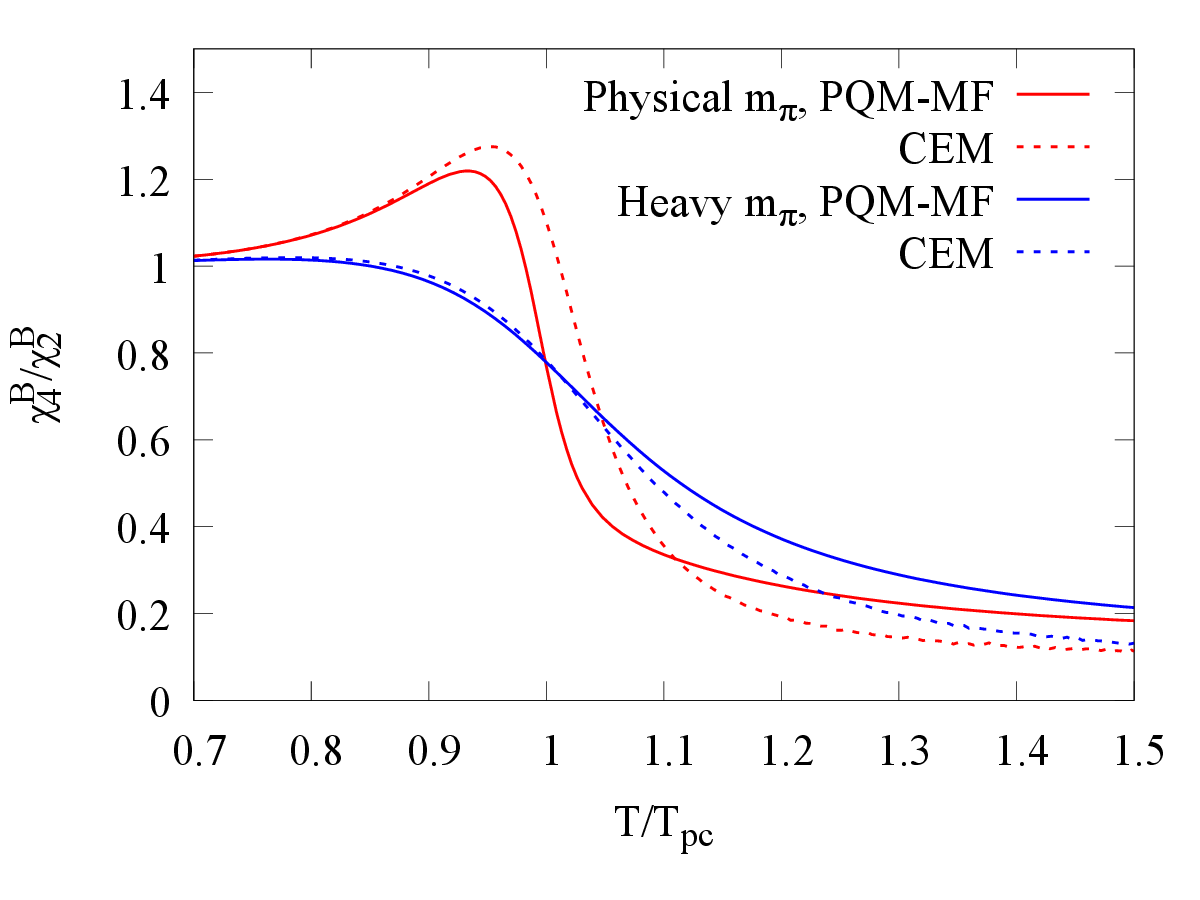}
 \includegraphics[width=0.32\textwidth]{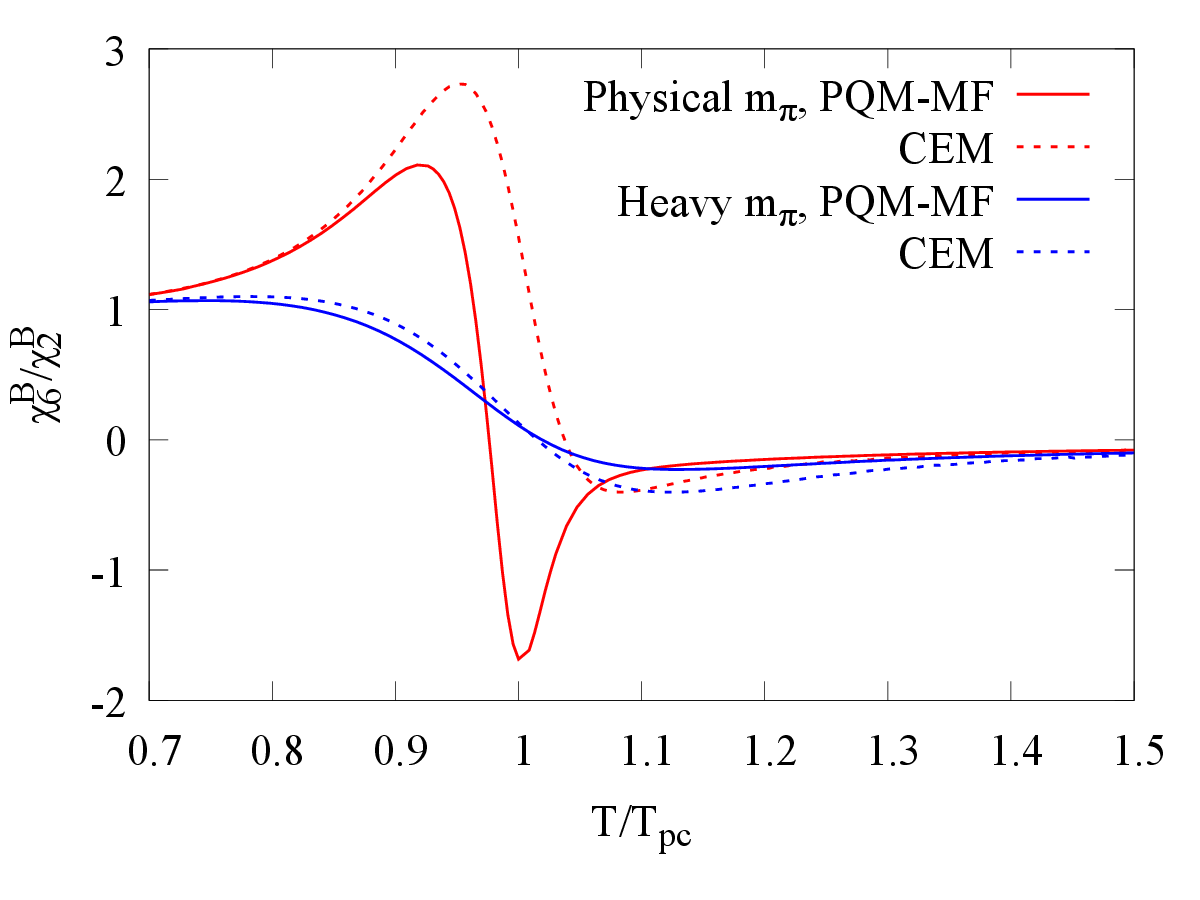}
 \includegraphics[width=0.32\textwidth]{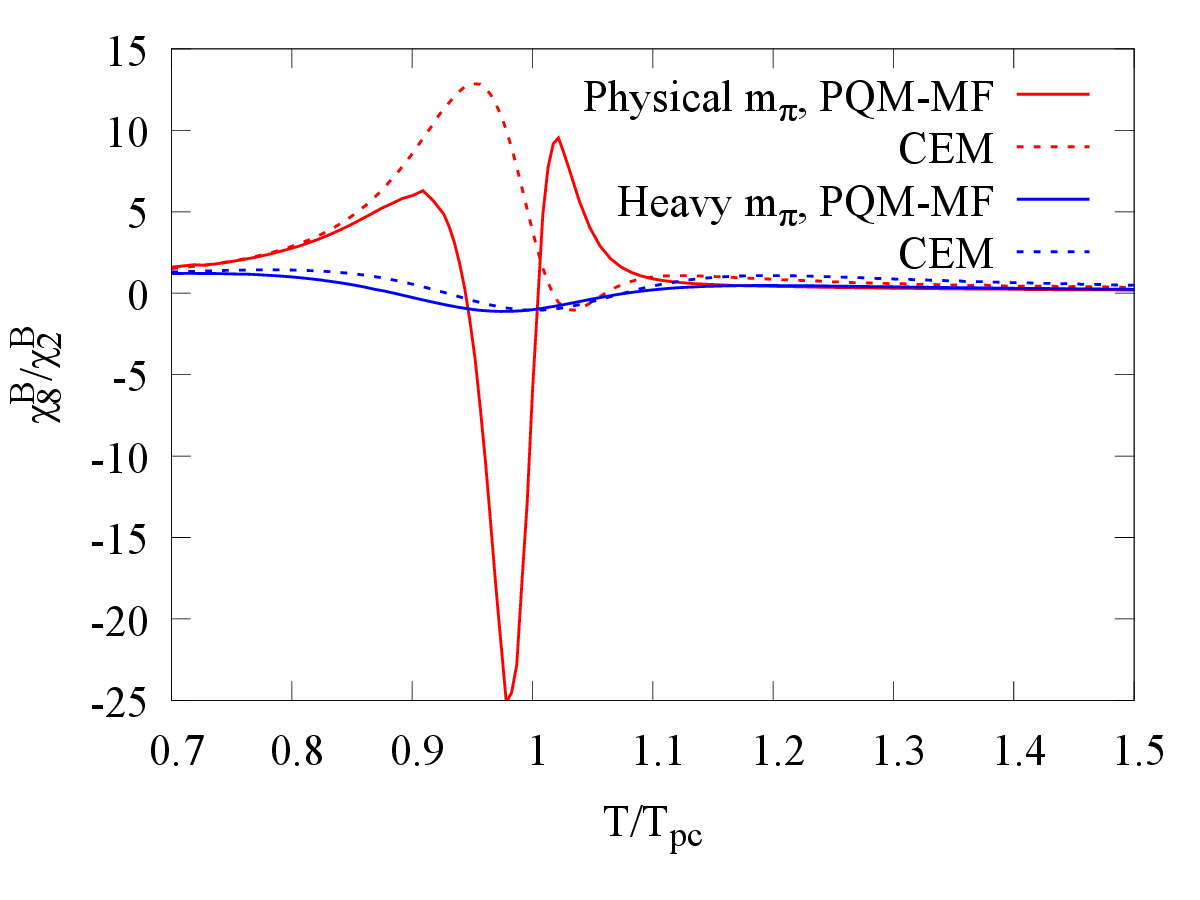}
 \caption{The ratio of the fourth- (left), sixth- (middle) and eighth-order (right) baryon number
 cumulants to the second-order one, calculated in the PQM model and using the CEM ansatz for different pion masses.  }
 \label{fig:cumulantratio}
\end{figure*}

\begin{figure*}[!tb]
 \centering
 \includegraphics[width=0.32\textwidth]{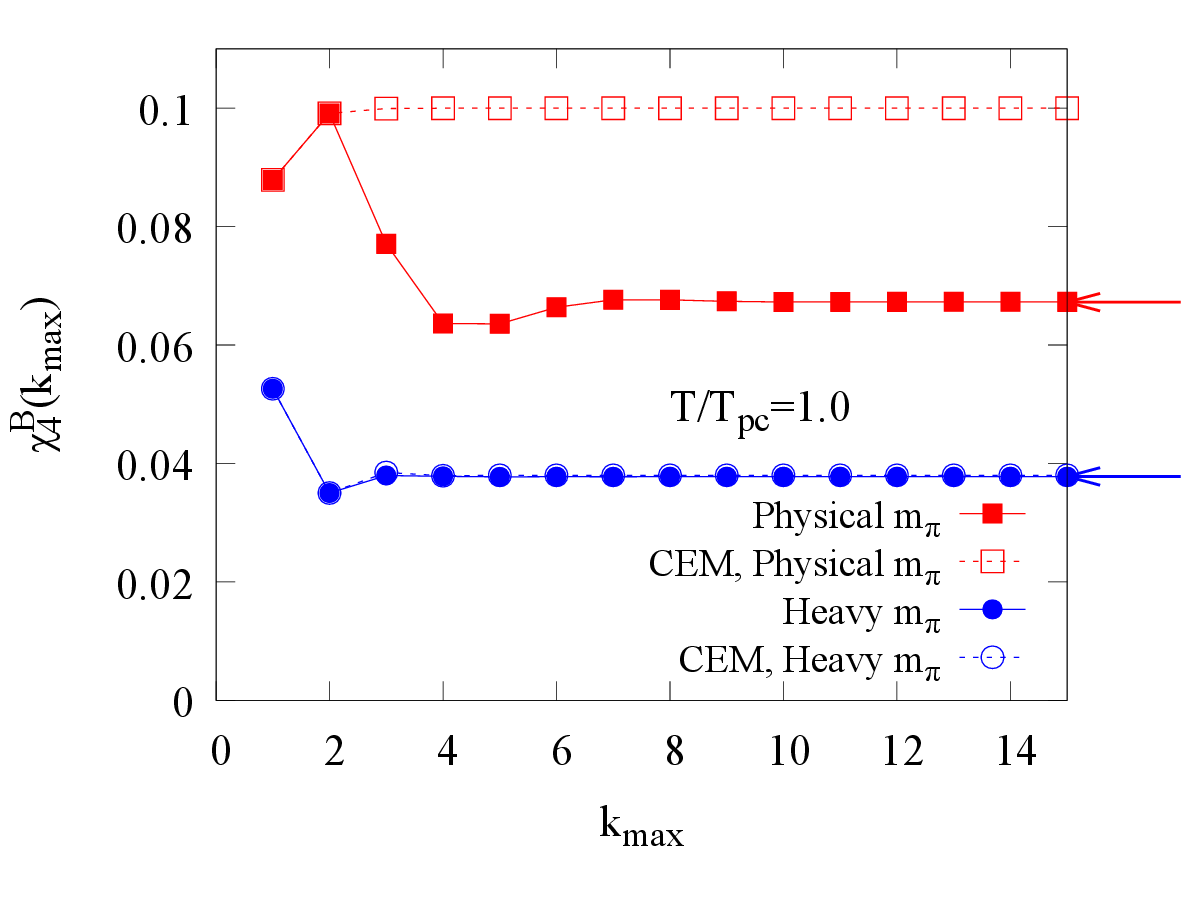}
 \includegraphics[width=0.32\textwidth]{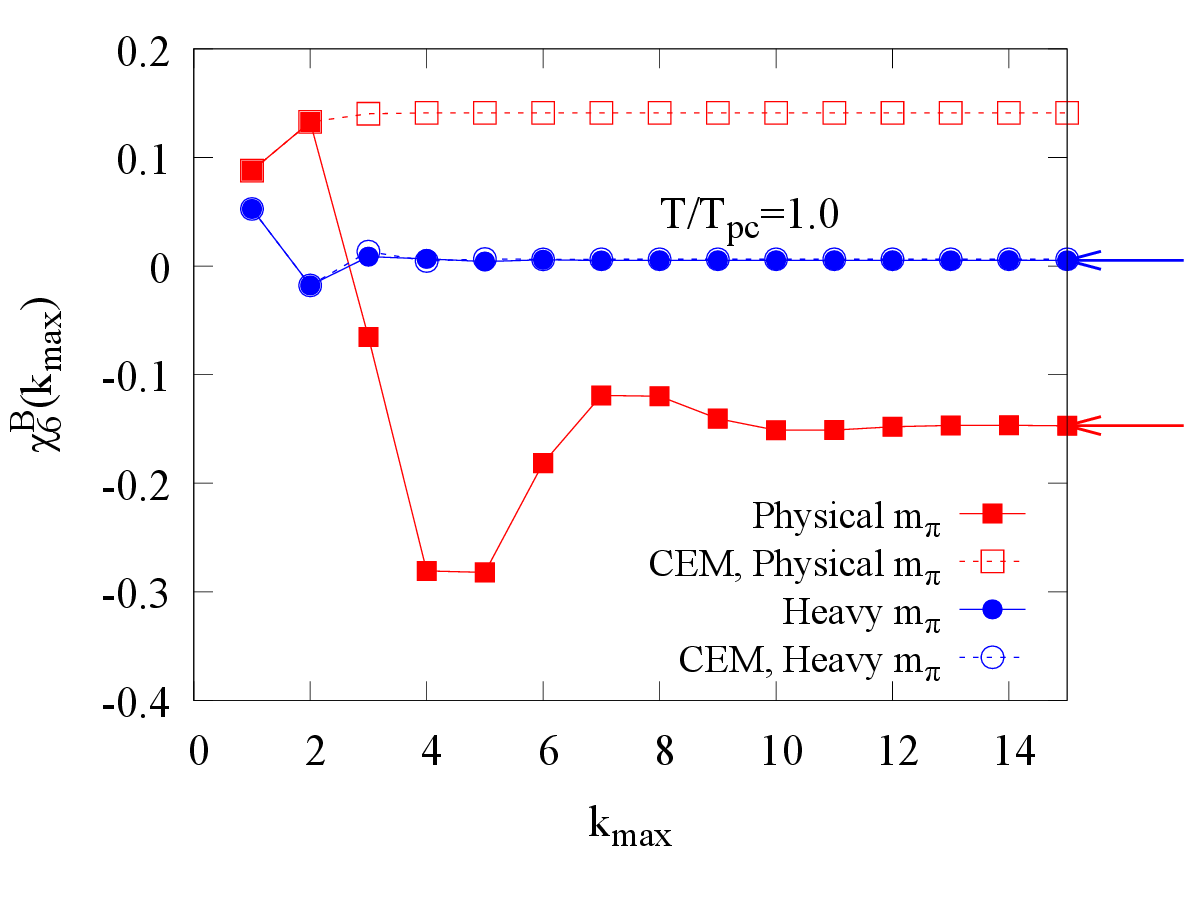}
 \includegraphics[width=0.32\textwidth]{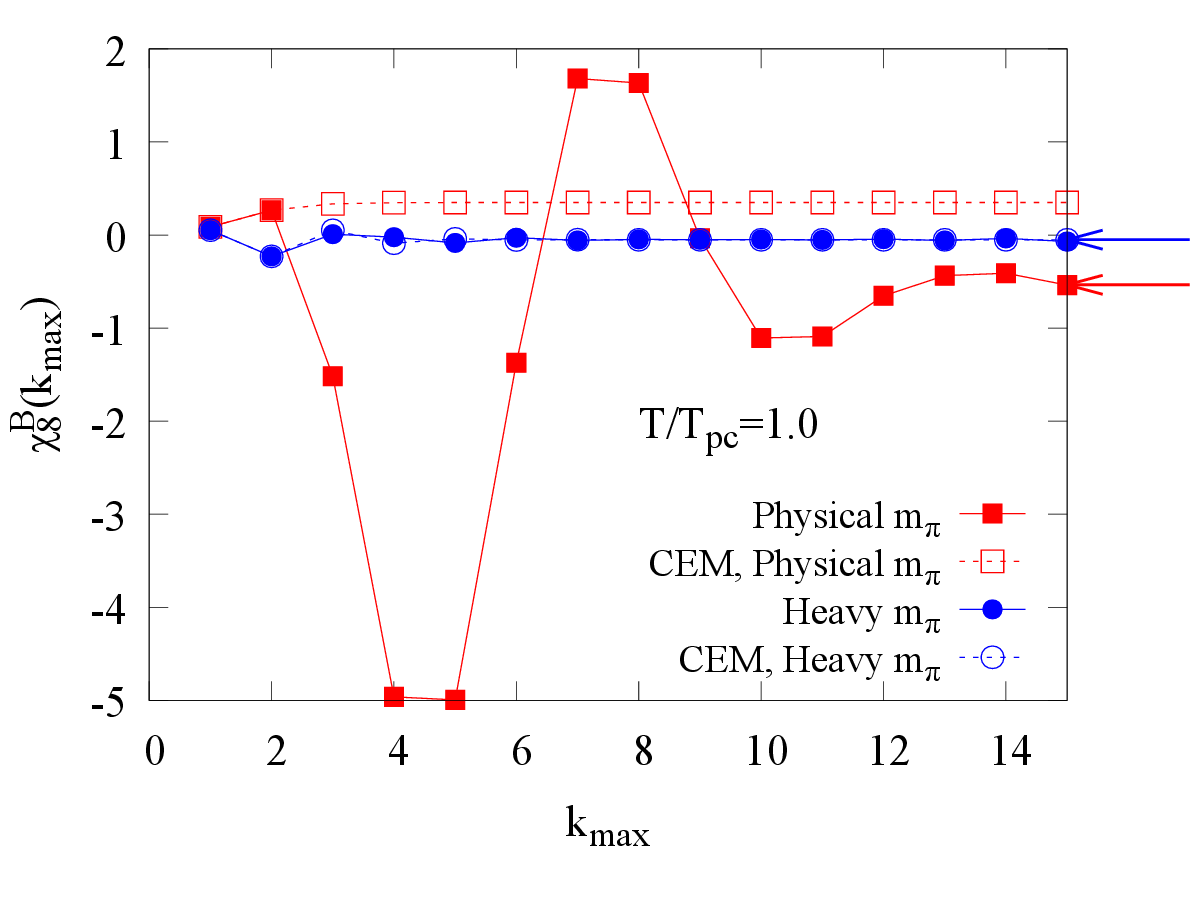}
 \caption{Fourth- (left),  sixth- (center), and eighth-order (right) 
 cumulants obtained using the partial sum up to $k=k_{\text{max}}$ [see Eq. \ref{eq:chin_sum}] in the PQM model  and adopting the CEM ansatz
  at $T=T_{pc}$. The arrows indicate the values of the
 cumulants obtained by taking derivatives of the pressure.}
 \label{fig:fluc_kmax}
\end{figure*}

We first compare the cumulant ratios obtained using the CEM scheme to those computed directly from derivatives of the  pressure. In the CEM scheme the summation in Eq. \eqref{eq:chin_sum} converges, due to the
exponential damping of higher-order terms, .

Figure~\ref{fig:cumulantratio} displays the
$\chi_4^B/\chi_2^B$, $\chi_6^B/\chi_2^B$, and $\chi_8^B/\chi_2^B$,  cumulant ratios.
As expected from the comparison of Fourier coefficients, these  cumulant ratios are rather well reproduced in the CEM scheme for a  heavy pion mass,  where criticality is strongly suppressed.
However, for a physical value of the pion mass, the CEM scheme does not reproduce the
$\chi_6^B/\chi_2^B$ and $\chi_8^B/\chi_2^B$ cumulant ratios, \footnote{Note that
    strictly speaking a mean-field calculation does not yield the
    correct critical behavior, which leads to divergences in $\chi_6^B$ and $\chi_8^B$ in the chiral limit. However, with a small explicit symmetry breaking term, the mean-field
    approximation yields baryon number cumulants that are in qualitative agreement with the expected behavior. }
which are sensitive probes of the chiral crossover transition
\cite{friman11:_fluct_as_probe_of_qcd}.
Although the CEM scheme roughly reproduces the sign structure of the cumulant
ratios, it clearly fails to capture the location of sign changes and the
magnitude of the fluctuations.

The reason for these differences between the cumulants of the PQM model  and those obtained in the CEM scheme can be traced back to the asymptotic behavior of the Fourier coefficients. In  Fig.~\ref{fig:fluc_kmax} we show the higher-order cumulants  computed  at $T=T_{pc}$.
Since each derivative with respect to the chemical potential brings one power of $k$ in  Eq.~\eqref{eq:chin_sum}, the higher-order cumulants  are
more sensitive to the higher-order coefficients.\footnote{We note that for cumulants, where the highest power of $k$ in the corresponding Fourier coefficient, $b_k k^{2n-1}$, is a non-negative integer, the Fourier series diverges due to $\delta$-functions or derivatives thereof at the integration boundaries, $\theta_B=\pm \pi$.}

In Fig.~\ref{fig:fluc_kmax} we show the  $\chi_n^B$  for $n=4, 6$ and $8$
as  functions of the upper limit of the summation
$k_{\text{max}}$ in Eq.~\eqref{eq:chin_sum}.
For a very heavy pion mass, there is almost no difference between $\chi_n^B$ computed using the Fourier coefficients of the PQM model and those obtained using the CEM scheme.
This is due to the rapid decrease of the higher-order coefficients $b_k$:
substantial differences between the CEM and the true Fourier coefficients appear  only at large
$k$. Since these are small, they do not contribute significantly to the cumulants, as indicated, e.g., by the rapid convergence of $\chi_8^B$, shown in Fig.~\ref{fig:fluc_kmax}.

There are, however, large differences between cumulants of the PQM model and those of the CEM scheme for a physical pion mass. This is due to the  higher-order coefficients, which in this case contribute substantially to the $\chi_n^B$. While the CEM cumulants converge rapidly with $k$, the convergence of the PQM  model results is much slower. In the latter, the asymptotic values of  $\chi_6^B$ and $\chi_8^B$ are, at $T=T_{pc}$, reached for $k_{\text{max}}=10$  and 15, respectively.

The partial sums for the sixth- and eighth-order cumulants shown in Fig.~\ref{fig:fluc_kmax}, clearly demonstrate that, for a physical $m_\pi$, the negative values of these cumulants at $T\simeq T_{pc}$ are by and large due to the large negative values of $b_3$ and $b_4$ (see the left panel of Fig.~\ref{fig:bk_tc}). With increasing temperature, the frequency of the oscillation increases, leading to the staggered sign structure and strong cancellations in the sum of Fourier coefficients in Eq.~\eqref{eq:chin_sum}. Consequently, at temperatures above $T_{pc}$, the cumulants $\chi_6^B$ and $\chi_8^B$ rapidly approach zero, as seen in Fig. \ref{fig:cumulantratio}.

\section{Concluding Remarks}
\label{sec:summary}

We have discussed the behavior of the Fourier coefficients of the
net baryon density and their relation to the singularities  in the complex chemical potential plane.

We found that the presence of singularities is reflected in the asymptotic behavior of the Fourier coefficients. This implies, e.g., that the existence or nonexistence of a phase transition cannot be settled by examining only a few low-order terms in the Fourier series.
{This was illustrated by considering the cluster expansion model \cite{Vovchenko:2017gkg}.  There,  the authors have introduced  the phenomenological  prescription
for computing the higher-order Fourier coefficients $b_n$
of the net baryon density  in terms
of the first two coefficients,  adopted  from lattice QCD  calculations.
Furthermore we have also introduced,  the model (FRM)  for  $b_n$  with $b_1$ and $b_2$ taken as input parameters  from LQCD.  In both  models, it was demonstrated   that knowledge of the first four $b_n$'s  is  not sufficient to draw any conclusion on  criticality. Moreover,  considering   the analytic structure of CEM  we have shown  explicitly that the Fourier series of this model  can be resummed and expressed by the polylogarithm functions which have well-defined analytic structure. In contrast to the expectation in Ref. \cite{Vovchenko:2017gkg}, this  excludes explicitly in CEM any information on  the QCD  phase transition.}

In order to explore  the influence of singularities associated with the
chiral phase transition on the Fourier coefficients, we employed the
PQM model as a low-energy effective approximation to QCD.

We then showed that the large-order
behavior of the Fourier coefficients exhibits characteristic oscillations
and exponential damping, due to the imaginary and real parts of the
chiral singularity, respectively. Consequently, in the chiral limit,
the Fourier coefficients at temperatures $T \leq T_c$  show only an exponential decay, while  at $T > T_c$ they exhibit a power-law decay superimposed on oscillations, whose frequency reflects the imaginary baryon chemical potential at the location of the critical point. As the pion mass is increased, the critical behavior is weakened because the singularity is removed from the imaginary axis and thus gains a larger real part, resulting in  a stronger damping of the Fourier coefficients.  Moreover, we have shown that both the Roberge-Weiss transition and the corresponding end point give rise to  characteristic power-law decays of the Fourier coefficients.

Based on the characteristics of the Fourier coefficients, we discussed the implications for the higher-order net baryon number cumulants. We pointed out that the signature of the chiral crossover transition in the net baryon number cumulants~\cite{friman11:_fluct_as_probe_of_qcd},  the sign change around $T_{pc}$, is related to the oscillatory behavior of  Fourier coefficients induced by the chiral branch-point singularity in
the complex chemical potential plane.

Our results indicate that the Fourier coefficients provide valuable information on the QCD phase transitions. Thus, lattice QCD calculations of these coefficients will improve our the understanding of the
phase structure of QCD. General results on the effect of critical singularities on the Fourier coefficients of the net baryon number
are reported in Ref.~\cite{Fourier_general}.

\section*{Acknowledgments}
We are grateful to Vladimir Skokov for useful conversations and thank Volodymyr Vovchenko for discussions on Ref. \cite{Vovchenko:2017gkg}, which inspired the  work reported here.
The work of G.A. was supported by the Hungarian OTKA Fund No. K109462 and
by the COST Action CA15213 THOR. K.M. acknowledges support from the RIKEN iTHES project and iTHEMS program. K.M. and K.R. acknowledge support by the Polish National Science Center NCN under Maestro Grant No. DEC-2013/10/A/ST2/00106, the Polish  Ministry of Science and Higher Education and by the Extreme Matter Institute EMMI at GSI. This work
was supported in part by the Deutsche
Forschungsgemeinschaft (DFG) through the
grant CRC-TR211 “Strong-interaction matter
under extreme conditions”.


\end{document}